\newcommand{\secref}[1]{Sec.~\ref{#1}}

\newcommand{\figref}[1]{Fig.~\ref{#1}}
\newcommand{\ud}{\mathrm{d}}	
\newcommand{\ie}{\emph{i.e. }} 
\newcommand{\eg}{\emph{e.g. }} 
\newcommand{\etal}{\emph{et~al. }} 

\documentclass[aip,graphicx,jcp,reprint]{revtex4-1}

\usepackage[version=3]{mhchem}
\usepackage[dvips]{epsfig}
\usepackage[latin1]{inputenc}
\usepackage{amssymb,mathrsfs,amsmath,amsfonts}
\usepackage{psfrag}
\usepackage{multirow,stackrel,dsfont,tabularx}

\usepackage{pstricks, overpic, pst-plot,pst-node}
\newsavebox{\IBox}

\graphicspath{{./figures_rev2/}{./}}

\begin{document}


\title{Model reduction for slow--fast stochastic systems with metastable behaviour}



\author{Maria Bruna}
\email{bruna@maths.ox.ac.uk}
\affiliation{Mathematical Institute, University of Oxford, Oxford, OX2 6GG, U.K.}
\affiliation{Computational Science Laboratory, Microsoft Research, Cambridge, CB1 2FB, U.K.}

\author{S. Jonathan Chapman}
\affiliation{Mathematical Institute, University of Oxford, Oxford, OX2 6GG, U.K.}

\author{Matthew J. Smith}
\affiliation{Computational Science Laboratory, Microsoft Research, Cambridge, CB1 2FB, U.K.}

\date{\today}

\begin{abstract}

The quasi-steady-state approximation (or stochastic averaging
principle) is a useful tool in the study of multiscale stochastic
systems, giving a practical method by which to reduce the number of
 degrees of freedom in a model. The method is extended here
 to slow--fast systems in which the fast variables exhibit metastable behaviour. The  key parameter that determines the form of the reduced model is the ratio of the timescale for the switching of the fast variables between metastable states to the timescale for the evolution of the slow  variables. The method is illustrated with two examples: one from biochemistry (a fast-species-mediated chemical switch  coupled to a slower-varying species), and one from ecology (a predator--prey system). Numerical simulations of each model reduction are compared with those of the full  system. 
\end{abstract}


\maketitle 

\section{Introduction} \label{sec:intro}

Understanding the impact of noise on nonlinear dynamical systems has
been an active field of research for many years, with a wide range
of applications in physics,  chemistry,  biology, ecology and
earth science. Although the addition of noise  sometimes does not change the qualitative dynamics  and can be modelled by adding a stochastic perturbation to the deterministic solution trajectory (as in the linear noise approximation \cite{Thomas:2012ej}), there is a growing number of applications in which noise has been shown to be crucial to explain features which cannot be captured by deterministic models.  For example, in deterministic systems with multiple stable steady states the addition of even a small amount of noise causes these states to become metastable, with the stochastic system undergoing random transitions between the deterministic steady states. Examples include genetic regulatory networks,\cite{McAdams:1999ex} lactose utilisation networks and the bet-hedging in bacteria.\cite{Ozbudak:2004gi,Veening:2013bj} This behaviour cannot be captured by the deterministic model, \cite{Horsthemke:1984ut} but can be explained successfully with a stochastic model.\cite{Wang:2012cd} 

\begin{figure*}[t]
\begin{center}
\includegraphics[width=.4\textwidth]{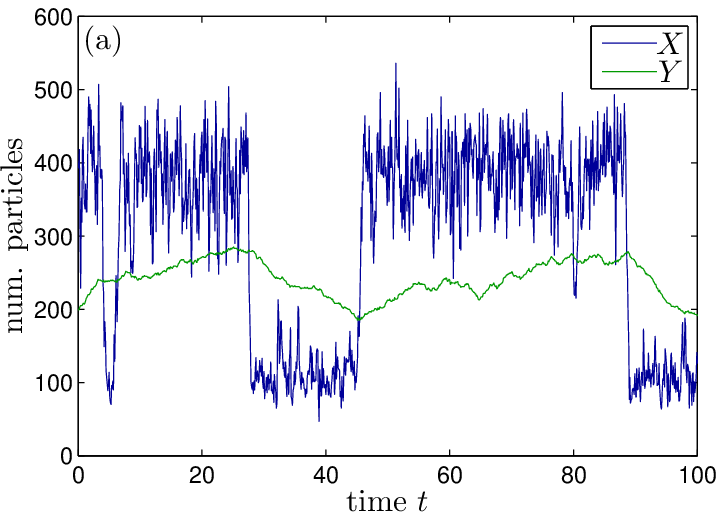} \quad 
\includegraphics[width=.4\textwidth]{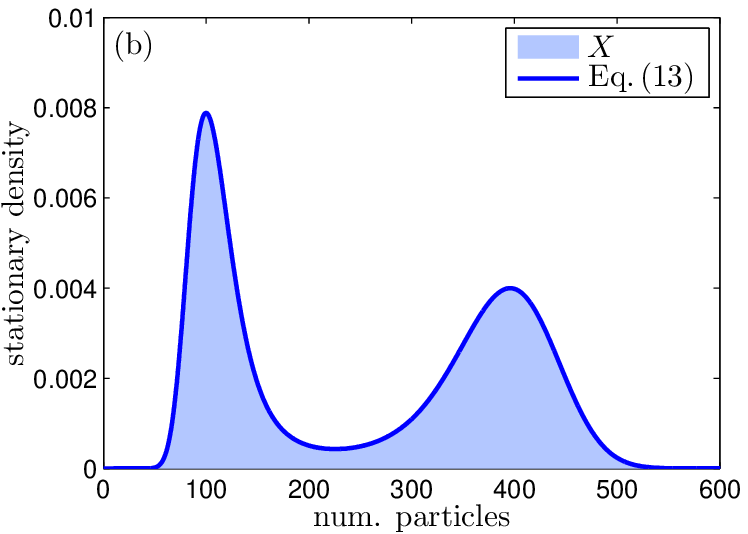} 
\caption{(a) Single realisation of \eqref{reactions} obtained using the exact NRM SSA with $X(0) = 400$ and $Y(0) = 200$. (b) Comparison of the stationary distribution $\rho (x)$ given by \eqref{stat_x} (solid dark blue line) and the histogram of $X$ from SSA of \eqref{reactions} ($10^7$ samples used, light blue area). We use $\tau_y = 25$ and the parameter values \eqref{values}.}
\label{fig:Fig1}
\end{center}
\end{figure*}

A common feature of many complex dynamical systems is the presence of processes evolving on widely separated timescales. This can present a challenge for numerical simulation. Often we are only interested in the behaviour of solutions on a long timescale,\cite{E:2005bq} but in principle to  determine this we need to resolve processes occurring on the fastest timescale. Metastability complicates this further, since the switching rate between states introduces a new implicit timescale to the process. For example, several ecological systems are believed to be able to switch between alternative states, such as between tree and grass-dominated vegetation.\cite{Staver:2011bz,Scheffer:2012ct} A few key mechanisms may be sufficient to explain the observed bi-stability of these systems, but a wide range of other ecosystem components (such as the resident bird community)  respond to these dynamics, and may do so at different rates; for example the population size of a resident bird community may respond relatively quickly to a sudden switch in the dominant vegetation type, but the carbon content of the soil may respond relatively slowly.   This raises questions about how to efficiently and accurately simulate the dynamics of such systems: can the evolution of the slowly evolving components be predicted without simulating the full evolution of all the fast components?   

In this paper we bring together these two features, multiple
timescales and metastability, and consider noisy systems with slow and
fast metastable components. We develop a method to remove the fast
degrees of freedom while retaining the metastable behaviour in
the resulting mathematical description of the slow processes.

Most of the work done to remove fast degrees of freedom from
stochastic multiscale systems is based on extending the principle of
slow or centre manifold theory of deterministic dynamical
systems. This assumes that the phase space can be decomposed into fast
variables $x$  which relax very quickly and slow variables $y$, which
change more slowly: 
\begin{subequations}
\label{general_system}
\begin{alignat}{2}
\epsilon \frac{\ud x}{\ud t} &= F(x, y),& \qquad x(0) &= x_0, \\
\frac{\ud y}{\ud t} &= G(x,y),& y(0) &= y_0, 
\end{alignat}
\end{subequations}
where $\epsilon \ll 1$ quantifies the separation of
timescales. The standard singular perturbation theory, based on
Fenichel's theory,\cite{Fenichel:1979dz} consists of  taking the
limit $\epsilon \to 0$ and assuming that the fast variables $x$ have
equilibrated onto the \emph{centre manifold}, $F(x,y) = 0$. The flow
on this slow manifold (with reduced dimensionality) is  then given by
\begin{equation}
\label{slow_deterministic}
\frac{\ud y}{\ud t} = G(h(y),y), \qquad  y(0) = y_0, 
\end{equation}
where $F(h(y), y) = 0$.\cite{Desroches:2012cj} Note that the
projection onto the slow manifold is valid even if $F(x, y) = 0$ has
more than one solution for $y$ fixed; given the initial condition
$x_0$, the fast system evolves deterministically to one of the
steady states as determined by $F(x,y)$. However, this is no longer true
when we introduce noise in the system: noise can make $x(t)$ fluctuate
between  several  states satisfying $F(x, y)  = 0$ and \eqref{slow_deterministic} does no longer capture the dynamics of $y(t)$. Developing a good approximation for the dynamics of system \eqref{general_system} in the presence of multiple stable states and noise is the key problem that we address here.

There exists a variety of approximation methods to generalise this
model reduction to stochastic systems. The theoretical foundation of
these can be traced back to the stochastic centre manifold theory
developed in Ref.~\onlinecite{Boxler:1989cb}. A good introduction to the existing
techniques, classified according to the stochastic system
representation, can be found in Ref.~\onlinecite{Constable:2013fu}. Consider the
following 
stochastic counterpart of \eqref{general_system}:
\begin{subequations}
\label{general_system_stoch}
\begin{align}
\label{general_system_stochx}
\ud X  &= \frac{1}{\epsilon} F(X, Y) \ud t + \frac{1}{\sqrt \epsilon} \sigma
(X,Y) \ud W_X(t), \ X(0) = X_0, \\ 
\ud Y &= G(X,Y) \ud t + \Sigma(X,Y) \ud W_Y(t) , \quad Y(0) = Y_0, 
\end{align}
\end{subequations}
where $W_X(t)$ and $W_Y(t)$ are independent standard Brownian motions. 
Even though the presence of noise makes rigorous analysis more
complicated, the idea is conceptually very simple: one assumes that,
``freezing'' the slow variables at $Y(t) = \hat y$, the fast process $X(t)$
reaches a unique stationary density $\rho^{\hat y} (x)$ (or rather a
\emph{quasi-steady density}), which is the analogue of the unique
steady-state in the deterministic system. The slow
stochastic system analogous to  \eqref{slow_deterministic} is then
obtained by averaging  in the fast variables over this density:
\begin{equation}
\label{dyintro}
\ud Y  = \overline G(Y) \ud t + \overline \Sigma(Y) \ud W_Y(t),
\end{equation}
where
\begin{equation}
\label{stochave}
\overline G(Y) = \! \int \! G(x,Y) \rho^{\hat y}(x) \, \ud x, \ \
\overline \Sigma(Y) = \! \int \! \Sigma(x,Y) \rho^{\hat y}(x) \, \ud x,
\end{equation}
a process known as \emph{stochastic averaging}.\cite{Wainrib:2012vj} This process is sometimes referred to as the
quasi-steady-state (QSS) reduction.\cite{gardiner2004handbook}
The changes in $Y(t)$ may push $X(t)$ away from its previous
quasi-steady density, but by assumption fast transients in $X(t)$ die
out quickly and a new quasi-steady density $\rho^{\hat y}(x)$
applies. Numerical schemes based on the QSS are analysed in Ref.~\onlinecite{E:2005bq}  and implementations of those can be found in Ref.~\onlinecite{Cao:2005gj} and \onlinecite{Rao:2003gu}.

The underlying assumption of the above method is that $X(t)$ is well
approximated by a random variable chosen from its steady-state
distribution $\rho^{\hat y}(x)$, conditioned on a fixed value of $Y(t) = \hat y$.\cite{Newby:2013gx} However, this might not be true if the fast
degrees of freedom exhibit metastability. 
Our goal in this paper is  to extend the model
reduction of slow--fast stochastic systems to scenarios in which the
fast variables do not have a single invariant measure over the timescales of
interest, but switch randomly between a number of invariant measures.

Consider the stochastic metastable process for a single variable $X(t)$. The standard example of
metastable behaviour is Brownian motion with a double well potential [resulting in a stationary density similar to the one shown in \figref{fig:Fig1}(b)]. On short timescales the particle is most likely found near one of the two
potential minima, but on long timescales the particle can transition over the
energy barrier that separates the two wells. This problem  itself can be
thought of a slow--fast system, where the fluctuations of $x(t)$ within a
well is the fast process, and the 
metastable transitions between wells
 play the role of a (discrete) slow process, since they typically
 occur on a timescale $\tau_s \gg \epsilon$. 

The aim of model reduction techniques for metastable processes is to
eliminate the fast degrees of freedom while retaining
the slow metastable transitions. The extension of the QSS to
metastability is known as quasi-stationary analysis (QSA) and is based
on WKB projection methods.\cite{Ward:1998ve} The main feature of model reduction in this context is that, instead of averaging over a global quasi-stationary density $\rho(x)$ (as in the QSS, which would average over all the metastable
basins), it is assumed that $X(t)$ relaxes in one basin of
attraction $B_j$ and is well-approximated by a stationary random
process chosen from the quasi-stationary density $\rho_{j}(x)$
restricted on basin $B_j$. Then the reduced low-dimensional model consists of a discrete jump process between states $j$, where the switching rates are calculated from the transition rates  of the original process $X(t)$.\cite{Hinch:2005eg,
  Newby:2012fz, Newby:2013gx} In higher-dimensional systems where analytical approximations are not possible, computational techniques can be applied to determine the state space of the metastable variables and to sample the transition probabilities. A recent survey of such techniques applied to the simulation of rare events in molecular dynamics (such as conformation changes) can be found in Ref.~\onlinecite{Hartmann:2013wi}. 

Now suppose $X$ is coupled to an additional slow variable $Y$.  For ease of exposition we focus on the simpler case in which
the dynamics of $X$ are 
independent of $Y$. As we will discuss
in section  \ref{sec:discussion}, the extension to the more general
case is conceptually straightforward.
We see that three timescales are required to characterise such a 
system: the timescale of the fast variables $\epsilon$, the timescale
of the \emph{switches} of the fast variables $\tau_s \gg \epsilon$, and the
timescale of the slow variables $\tau_y \gg \epsilon$. We are interested in
approximating the system on the slowest of these timescales.

 We will examine how the reduced model changes with the
relative size of  $\tau_s$ and  $\tau_y$. If $\tau_y  \ll \tau_s$, the reduced model is simply a
discrete jump process.\cite{Hinch:2005eg} If $\tau_y
\gg \tau_s$, the slow variables evolve on a timescale which is even
slower than the switches in $x$ and the classical QSS approximation
applies. In the intermediate regime in which $\tau_y = \mathcal
O(\tau_s)$ things become more interesting, as the evolution of 
 $y$ critically depends on the metastable
behaviour of $x$. This is the main contribution of the present work.  

The rest of this paper is organised as follows. In Sec. \ref{sec:modelproblem} we
introduce a model chemical system with which we will illustrate the model reduction method.
 In section \ref{sec:intuitive} we describe the idea 
behind our analysis and show how to obtain a reduced model in the slow
timescale when the fast process is monostable, and how this changes
when there is metastability. We  introduce the three characteristic
timescales in our slow--fast model system with bistability, from which
we define three regimes depending on their relative sizes. We present
the three corresponding reduced models and show numerical comparisons between
these reduced slow-scale stochastic models and the full slow--fast
process using a stochastic simulation algorithm (SSA).  
In section \ref{sec:modelproblem2} we apply the method to a simple
predator--prey model with bistable prey and use it to estimate the mean
extinction time of the predator. More background on why we choose this particular case study will be given in the beginning of that section.  
In the concluding section \ref{sec:discussion} we briefly discuss the
results in the context of current directions of research in
theoretical ecology and stochastic nonlinear dynamics. The justification of the reduced models through a perturbation analysis of the corresponding Fokker--Planck equations is presented in appendix
\ref{sec:perturbation}.

\section{Model problem}
\label{sec:modelproblem}

We introduce the following model  as an example of a slow--fast
system with bistability.  
A system with two chemical species $X$ and $Y$ changes according to
the  reactions  
\begin{subequations}
\label{reactions}
\begin{gather}
\label{reactions_x}
\ce{$\emptyset$ <=>[k_1/\epsilon][k_2/\epsilon] $X$}, \hspace{2cm}  
\ce{$2X$  <=>[k_3/\epsilon][k_4/\epsilon] $3X$}, \\
\label{reactions_y}
\ce{$\emptyset$  <=>[k_5(X)/\tau_y][k_6/\tau_y] $Y$}
\end{gather}
\end{subequations}
Note that the evolution of the chemical $X$ is self-determining, while the
production of chemical $Y$ depends on the amount of $X$ present in the
system through the function $k_5(x)$. We suppose the system is in a
well-mixed reactor of unit  volume. 
We have chosen to write the reaction rates for $X(t)$ in 
terms of fixed $k_1$ to $k_4$ and a small parameter $\epsilon$; 
by varying $\epsilon$ we can change the timescale for the
evolution of $X$ without changing the equilibrium distribution. 
Similarly, we have chosen to write the reaction rates for 
$Y(t)$ in terms of fixed $k_5(x)$ and $k_6$ and a parameter
$\tau_y$.

The set of four reactions \eqref{reactions_x} governing $X(t)$ is
known as the  Schl\"ogl model.\cite{Schlogl:1972he} This model has two favourable states (or basins of attraction) for some parameter values, and the stochastic system displays bistable switching between these states even for parameter values for which the deterministic model is monostable.\cite{Erban:2009ew}

The stochastic model of the six chemical reactions \eqref{reactions}
introduces the propensity functions 
 \begin{alignat*}{2}
\alpha_1 (x) &= k_1/\epsilon, & \qquad \alpha_2(x) &= k_2 x/\epsilon, \\
\alpha_3(x) &= k_3 x (x-1)/\epsilon, &
\alpha_4(x) & = k_4 x (x-1) (x-2)/\epsilon, \\
\alpha_5 (x) & =
k_5(x)/\tau_y,&  \alpha_6(y) &= k_6 y/\tau_y, 
\end{alignat*}
which are such that the probability of reaction $i$ happening in time
${\rm d} t$ is $\alpha_i {\rm d} t$.\footnote{Some authors would use the convention that the propensity function for the third reaction, 
$2 X \stackrel[]{k_3/\epsilon}{\longrightarrow} 3X$, is $k_3 x (x-1)/2\epsilon$ instead
  of $k_3 x (x-1)/\epsilon$. See Ref.~\onlinecite{Gillespie:1977dc} for a discussion on
  conventions regarding reaction rates.} 
The system may then be simulated by using a stochastic
simulation algorithm (SSA) such as the exact Gillespie's Direct Method.\cite{Gillespie:1977dc} Here we use the equivalent exact and efficient Next Reaction Method (NRM).\cite{Gibson:2000jq}

The timescale for the evolution of $X$ is set by the decay rate 
$k_2/\epsilon$.
Similarly the
timescale for the evolution of $Y$ is set by the decay rate  $k_6/\tau_y$. 
Thus if $\epsilon$ and $\tau_y$ are to represent these timescales we
should choose $k_2$ and $k_6$ to be of order unity. The other
rates determine the equilibrium values of $X$ and $Y$. If we suppose a typical
equilibrium value of $X$ is  $O(\delta^{-1})$, this corresponds to 
$k_1=O(\delta^{-1})$, $k_3 = O(\delta)$, $k_4 = O(\delta^2)$. In \figref{fig:Fig1}(a) we plot one realisation of the SSA of \eqref{reactions} obtained for $X(0)=400$, $Y(0) = 200$, $\tau_y = 25$, and
\begin{equation}
\label{values}
\begin{aligned}
k_1 &=  0.6\, \delta^{-1}, & \quad  k_2 &=  1, &\quad k_3 &=  0.48 \,\delta, \\
k_4 &= 0.0666\,\delta^2, & 
k_5(x) &= 0.833 \,x, & k_6&= 1,\\
\epsilon&= 0.025, & \delta&=0.01. &&
\end{aligned}
\end{equation} 
The parameters values are chosen such that system \eqref{reactions} is bistable in $X(t)$ (as in Ref.~\onlinecite{Erban:2009ew}) and $Y(t)$ evolves in the same timescale as switches in $X$. The rate $k_5(x)$ (which we will also use in the simulations that follow) is such that the production reaction for $Y$ can be written as $X \longrightarrow X+Y$ with rate $0.833/\tau_y$. We also plot the stationary distribution of $X$, obtained by a long time simulation of the SSA with final time $t= 10^6$ and recording the value of $X$ every $\Delta t = 0.1$ in \figref{fig:Fig1}(b). We see that for these parameter values the system \eqref{reactions}
has two favourable states for $X$.  We see that $X(t)$ spends quite a long time fluctuating around one of these states before switching to the other. In section A 2 we will derive an analytical approximation of the mean switching times between states and find that is exponentially large in the parameter $\delta$. 
We want to understand how the oscillation between these two
states influences the dynamics of $Y$, and obtain an efficient and
accurate way to take this effect into account without 
having to simulate the full $X$ dynamics. 

\subsection{Continuous approximation: the chemical Langevin and the Fokker--Planck equations}
\label{sec:continuous}

For ease of exposition we will apply our model reduction techniques
not to \eqref{reactions} directly, but to the
continuous approximation of this system given by the chemical Langevin
equation. This approximation is similar to the numerical method of
$\tau$-leaping.\cite{Cao:2006hq} The idea of the approximation is
that for large $X$ 
and $Y$ the propensity functions do not change significantly after each
individual reaction event,  which enables us to jump forward many
reaction events without updating the propensities. The change in the number of molecules  between timesteps is then Poisson distributed.
If we now take a limit in which many reactions are considered between timesteps, but still with a small relative change in the number of molecules, $X$ and $Y$ may be approximated by continuous random variables and the change in molecular numbers between timesteps is approximately normally distributed (see Refs.~\onlinecite{Erban:2009ew,Gillespie:2000ig} for more details of this approximation). In that case we arrive at the chemical Langevin equations
\begin{subequations}
\label{sde_xy}
\begin{align}
\label{sde_x}
\ud X  &= \frac{v(X)}{\epsilon} \ud t + 
\sqrt{\frac{2 d(X)}{\epsilon}}\, \ud W_X(t), \\ 
\label{sde_y}
\ud Y &= \frac{V(X,Y)}{\tau_y} \ud t +
\sqrt{\frac{2 D(X,Y)}{\tau_y} }\, \ud W_Y(t), 
\end{align}
\end{subequations}
which are of the same form as the slow--fast system of stochastic differential equations \eqref{general_system_stoch}. The drift coefficients $v(x)$, $V(x,y)$ and the diffusion coefficients $d(x)$ and $D(x,y)$ are given by
\begin{align}
\nonumber v(x) & = k_1 - k_2 x + k_3 x(x-1) - k_4 x (x-1) (x-2),\\
\nonumber d(x) & = \frac{1}{2} \! \left[ k_1 + k_2 x + k_3 x(x\!-\!1) + k_4 x (x-1) (x-2) \right],\\ 
\nonumber V(x,y) & = k_5 (x) - k_6 y,\\
\label{drift_and_diffusion}
D(x,y) & = \frac{1}{2} [k_5 (x) + k_6 y].
\end{align}
Let $P(x, y, t) \ud x \ud y$ be the probability that $X(t) \in [x, x + \ud x)$ and $Y(t) \in [y, y+\ud y)$ at time $t$. Then the chemical Fokker--Planck (FP) equation for the joint probability distribution function corresponding to \eqref{sde_xy} is given by
\begin{equation}
\label{fokker-planck}
\begin{aligned}
\frac{\partial P}{\partial t} (x,y,t) = &
\frac{1}{\epsilon}\frac{\partial}{\partial x} 
\left\{  \frac{\partial}{\partial x} [ d(x) P] - v(x) P \right\}\\
& +
\frac{1}{\tau_y}\frac{\partial}{\partial y} \left\{  \frac{\partial}{\partial y} [
  D(x,y) P] - V(x,y) P \right\}. 
\end{aligned}
\end{equation}
This equation is complemented with no-flux boundary conditions on $[0, \infty)^2$ since the probability $P$ must remain normalised for all times. 

The stationary distributions of $X$ and $Y$, $\rho(x)$ and $q(y)$
respectively, correspond to taking the limit $t\to
\infty$ in the marginal probability densities 
\begin{equation}
\label{marginals}
\rho (x) = \lim_{t\to \infty} \int P(x,y,t) \, \ud y, \ \ q(y) = \lim_{t\to \infty} \int P(x,y,t)\, \ud x.
\end{equation}
Since the dynamics of $X$ do not depend on $Y$, it is
straightforward to solve for $\rho(x)$ by integrating
\eqref{fokker-planck} with respect to $y$ to give 
\begin{equation}
\label{fokker-planck_x}
 \frac{\partial}{\partial x} \left ( \frac{\partial}{\partial x} [ d(x) \rho ] - v(x) \rho  \right )=0,
\end{equation}
where we have used no-flux boundary conditions at $y = 0$ and $y=
\infty$. Thus the stationary distribution of $X$ is given by  
\begin{equation}
\label{stat_x}
\rho(x) = \frac{A}{d(x)} \exp \left [ \int_0^x \frac{v(s)}{d(s)} \ud s \right],
\end{equation}
where $A$ is the normalisation constant. The solution \eqref{stat_x} is plotted in  \figref{fig:Fig1}(b) as a blue solid line. As expected, the results compare well with the results obtained by the long time stochastic simulations.

\section{Reduced slow-scale models via stochastic averaging}
\label{sec:intuitive}

In this section we show how the stochastic model \eqref{reactions} may
be reduced when $\epsilon$ is small, and how the reduced model depends on
the size of $\tau_y$ relative to the size of $\tau_s$.

\subsection{The case when $X(t)$ is
  monostable} \label{sec:Xmonostable} 

\begin{figure*}[t]
\begin{center}
\includegraphics[height=.28\textwidth]{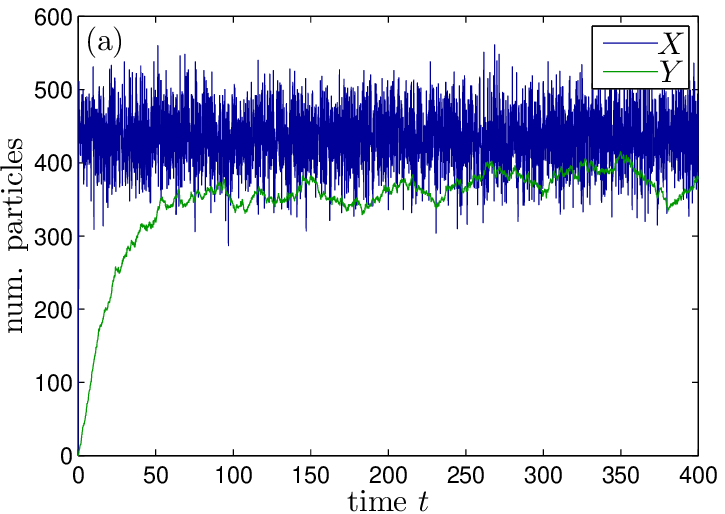} \quad
\includegraphics[height=.28\textwidth]{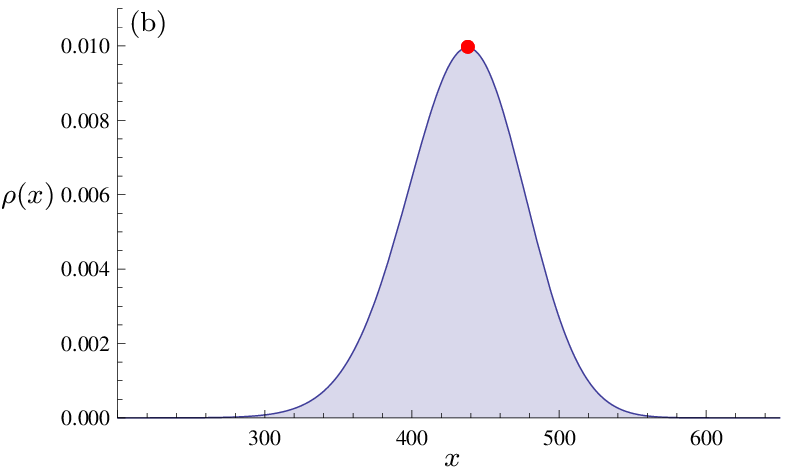} 
\caption{(a) Single realisation of \eqref{reactions} obtained using the exact NRM SSA. (b) Marginal stationary distribution $\rho (x)$ given by \eqref{stat_x} with the stable point $x_0$ marked with the red circle. We use $\tau_y = 25$, $X(0) = Y(0) = 0$, and the parameter values \eqref{values} except that $k_1 = 0.8 \delta^{-1}$.}
\label{fig:Fig2}
\end{center}
\end{figure*}

To give some context to the reduced models which follow, we first
consider the simpler case in which the fast species $X$ is
monostable. 
In \figref{fig:Fig2}(a) we present a
stochastic simulation of \eqref{reactions} with $X(0) = Y(0) = 0$ for $\delta = 0.01$, $\epsilon=0.025$, $\tau_y=25$, $k_1 = 0.8 \delta^{-1}$, and $k_2$ to $k_6$ as in \eqref{values}. 
In contrast to \figref{fig:Fig1}, we see that for this value of
$k_1$ the stochastic model fluctuates 
about a single value. This is demonstrated in
\figref{fig:Fig2}(b), where we plot the stationary
distribution $\rho(x)$ using \eqref{stat_x}. The maximum of the
stationary distribution is $x_0 \approx 437.8$, which satisfies
$v(x_0) - d'(x_0) = 0$.

In this case, since the relaxation time of $X$ to
its equilibrium density is much faster than the decay time of the slow
reaction,  the slow chemical reactions
\eqref{reactions_y} can be well-approximated by\cite{Cao:2005gj,Rao:2003gu} 
\begin{equation}
\label{reactiony_mono}
\ce{$\emptyset$  <=>[{\overline k_5}/\tau_y][k_6/\tau_y] $Y$}
\end{equation}
where $\overline k_5$ is the average production rate, given by
\begin{equation}
\label{k5eff}
\overline k_5 = \int_0^\infty k_5 (x) \rho(x) \, \ud x.
\end{equation}
For our particular example in which  $k_5(x) = k_5 x$, the effective
production rate is simply $\overline k_5 = k_5  \overline{X}$.

The analogous quasi-steady-state reduction in the 
chemical Langevin equation \eqref{sde_y} is 
\[
 \ud Y=  \frac{\overline V(Y)}{\tau_y}\, \ud t +  \sqrt{ \frac{2 \overline
   D(Y)}{\tau_y}} \,
 \ud W_Y(t), 
\]
where
\[
\overline V(y) = \int V(x, y) \rho(x) \, \ud x, \qquad \overline D(y) = \int D(x,y) \rho(x) \, \ud x.
\]
like in the stochastic averaged model \eqref{dyintro} and \eqref{stochave} described in \secref{sec:intro}.

\subsection{The case when $X(t)$ is bistable}  \label{sec:Xbistable}

\begin{figure}[bh]
\begin{center}
\includegraphics[height=.275\textwidth]{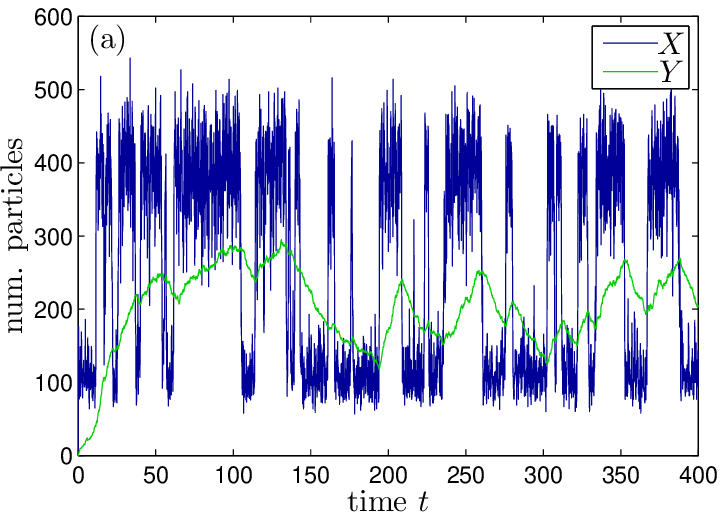} \\
\includegraphics[height=.275\textwidth]{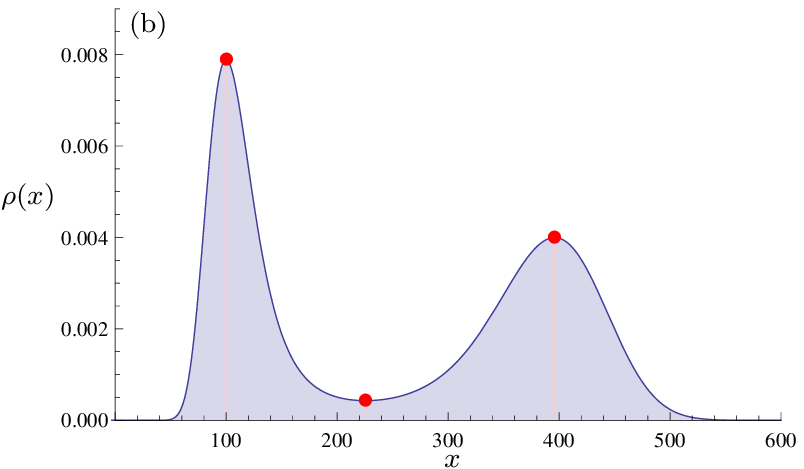} 
\caption{(a) Single realisation of \eqref{reactions} obtained using the exact NRM SSA. (b) Marginal stationary distribution $\rho (x)$ given by \eqref{stat_x} with the two stable fixed points $x_\pm$ and unstable fixed point $x_*$ marked with red circles. We use $\tau_y = 25$, $X(0) = Y(0) = 0$, and the parameter values \eqref{values}.}
\label{fig:Fig3}
\end{center}
\end{figure}
\begin{figure*}[tp]
\begin{center} 
\includegraphics[width=.32\textwidth]{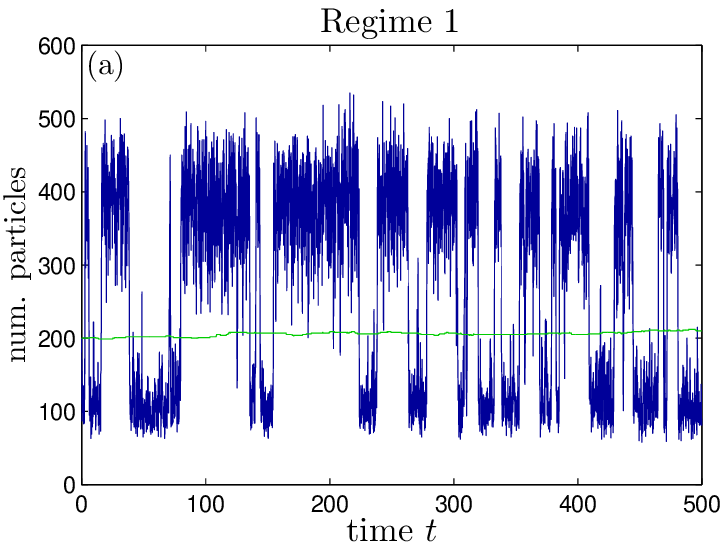}  \
\includegraphics[width=.32\textwidth]{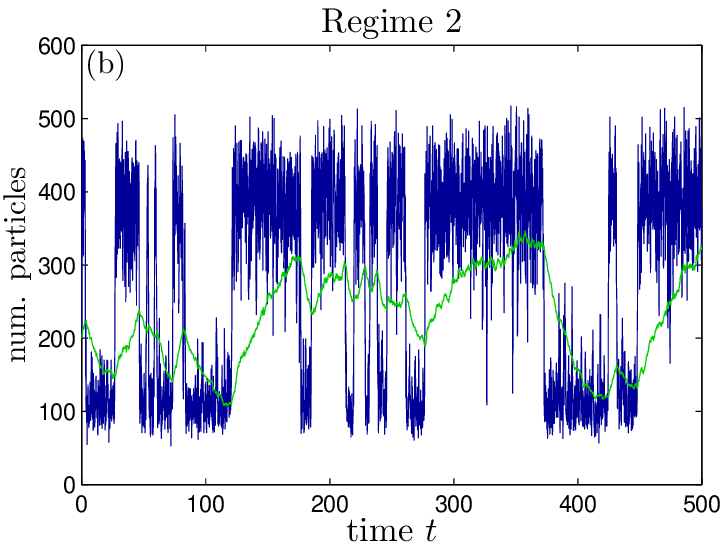} \
\includegraphics[width=.32\textwidth]{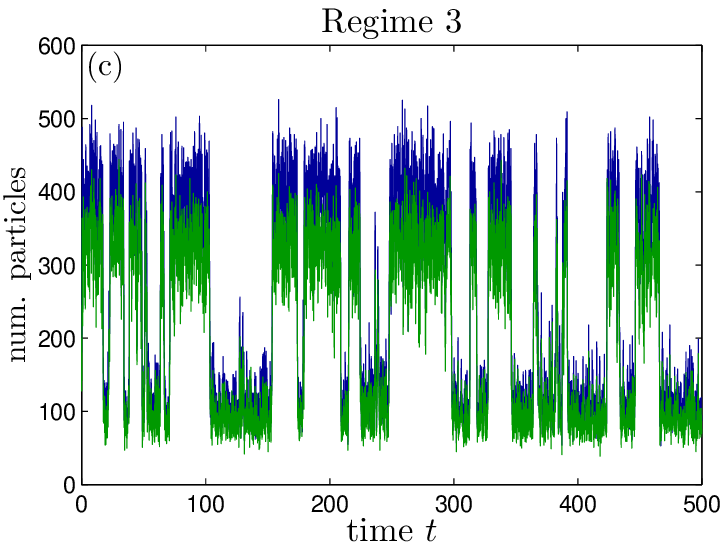} 
\caption{Evolution of $X(t)$ (blue lines) and $Y(t)$ (green lines) from a single realisation of \eqref{reactions} obtained using the exact SSA. We use the parameter values \eqref{values}, $X(0) = Y(0) = 200$, and vary $\tau_y$ to change between regimes.  (a) Regime 1: $\tau_y = 2500$. (b) Regime 2: $\tau_y = 25$. (c) Regime 3: $  \tau_y = 0.25$. }
\label{fig:Fig4}
\end{center}
\end{figure*}

Let us now consider the case when $X(t)$ is bistable, that is,
it switches between two favourable states as shown in
\figref{fig:Fig1}. Figure \ref{fig:Fig3}(a) shows an
illustrative trajectory of the system \eqref{reactions} for $X(0) =
Y(0) = 0$, with parameter values given by \eqref{values}. 
Figure \ref{fig:Fig3}(b) shows the stationary density $\rho(x)$
computed using \eqref{stat_x}. We denote the two peaks of this density by $x_\pm$ and the relative minimum (or unstable node) as $x_*$.  Now it is not clear whether it is appropriate to use the standard stochastic averaging technique, since  the assumption that $X(t)$ converges quickly to a stationary process with measure $\rho(x)$ is challenged by the metastable  behaviour of $X(t)$.

 To obtain a reduced slow system now we first need to characterise this
 metastable behaviour. In particular, we see that a third
 timescale $\tau_s$ emerges in the problem as the characteristic time
 for switches of the fast variable $X(t)$ between the two
 favourable states. Typically this timescale is much longer than the
 relaxation timescale for $X(t)$ within each well (indeed, this is the
 definition of metastability used here).

Thus we have the following scenario. On a short timescale  
 $X(t)$ relaxes to a quasi-stationary distribution centred around one
 of $x_-$ or $x_+$. On the longer timescale of $\tau_s$ the system
 switches from one of these distributions to the other, as $X(t)$ makes
 the transition between wells.
The nature of the reduced mode depends crucially on the relative sizes
of the switching timescale $\tau_s$ and the timescale for the slow
process $\tau_y$. We will see that there are three parameter regimes,
corresponding to $\tau_s \ll \tau_y$, $\tau_s \sim \tau_y$ and $\tau_s
\gg \tau_y$, respectively. Illustrative simulations of each of these
regimes are shown in \figref{fig:Fig4}.
In Subsections \ref{sec:rm1}--\ref{sec:rm3} we present the reduced stochastic
model appropriate for each regime, together stochastic
simulations comparing the reduced  model to the full system.
The mathematical justification for the reduced models is given in appendix \ref{sec:perturbation}.

\subsection{Regime 1:  $\epsilon\ll \tau_s \ll \tau_y$} \label{sec:rm1}
A sample  trajectory in this regime is shown in
Figure \ref{fig:Fig4}(a).   
In this case, $X$ has time to fully equilibrate on the timescale
of the evolution of $Y$. 
Thus the standard stochastic averaging can be used,
and the effective production rate of $Y$ can be computed as if
$X$ was monostable as 
\begin{equation}
\label{reactiony_reg1}
\ce{$\emptyset$  <=>[{\overline k_5}/\tau_y][k_6/\tau_y] $Y$}, \qquad
\overline k_5 = \int_0^\infty k_5 (x) \rho(x) \, \ud x.
\end{equation}
We call \eqref{reactiony_reg1} \emph{reduced model 1} or RM1. A derivation of this model using the associated Fokker--Planck equation is given in \ref{sec:regime1}.

\begin{figure}[bh]
\begin{center} 
\includegraphics[width=\columnwidth]{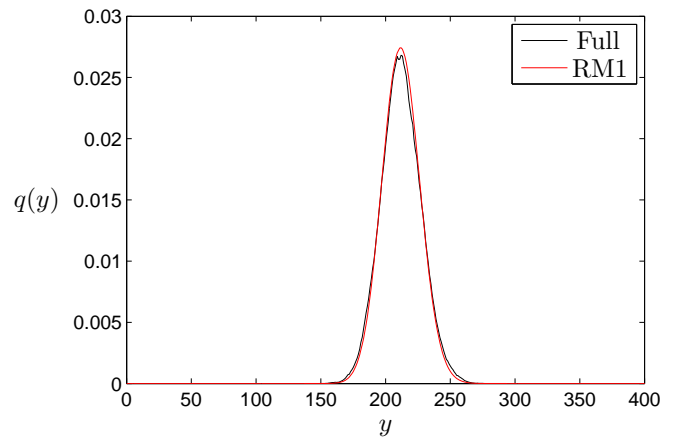} 
\caption{Comparison of the marginal stationary density $q(y)$ in regime 1: full system \eqref{reactions} (solid black line) and approximate reduced model 1 \eqref{reactiony_reg1} (dash red line). Both histograms are computed by running the NRM algorithm up to $t = 10^7$ taking recordings every $\Delta t = 5$ ($2\times 10^6$ samples). We use $\tau_y = 2500$  and the parameter values \eqref{values}.}
\label{fig:Fig5}
\end{center}
\end{figure}

\begin{figure}[hbt]
\begin{center}
\includegraphics[width=.95\columnwidth]{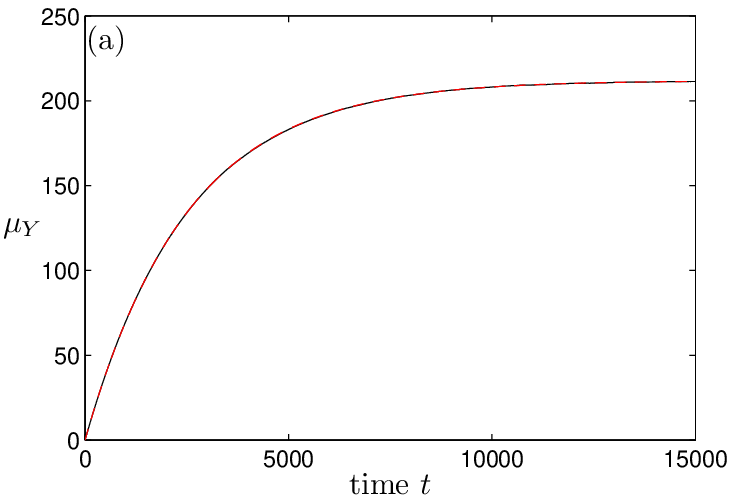} \\
\includegraphics[width=.95\columnwidth]{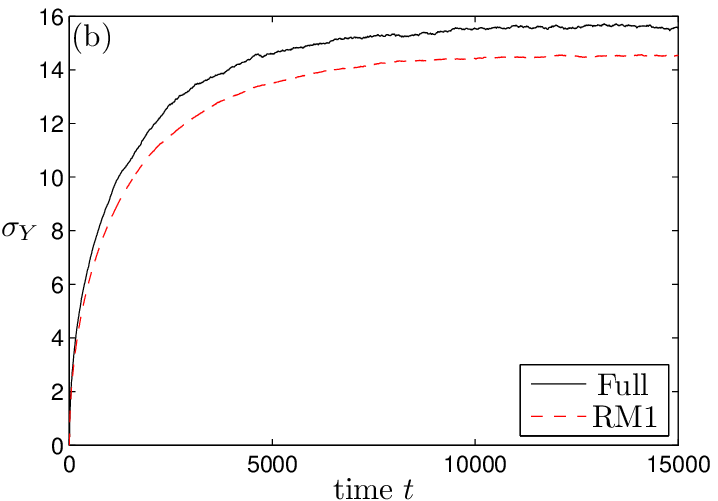}
\caption{Comparison of the time-dependent (a) mean $\mu_Y(t)$ and  (b) standard deviation $\sigma_Y(t)$: full system \eqref{reactions} (solid black line) and approximate reduced model 1 \eqref{reactiony_reg1} (dash red line). Black and red curves are computed as the average over $1.1\times 10^4$ and $10^5$ realisations respectively with initial conditions $Y(0) = 0$ and $X(0) \sim \rho(x)$ (for \eqref{reactions} only). We use $\tau_y = 2500$ and the parameter values in \eqref{values}. }
\label{fig:Fig6}
\end{center}
\end{figure}

In Figures \ref{fig:Fig5} and \ref{fig:Fig6} we compare simulation results (using the NRM SSA) of the full system \eqref{reactions} and the reduced system RM1. We use a long timescale for $Y(t)$, $\tau_y = 2500$, so that the system is in Regime 1. The stationary marginal density $q(y)$ (histograms obtained by a long time simulation of both models) is plotted in \figref{fig:Fig5}. Clearly, the reduced model provides a very good approximation to the histogram coming from the full-system simulation. The benefit of eliminating the fast variable is illustrated by the fact that the RM1 histogram took only 0.8 seconds to compute while the full system histogram required over 14 hours of computing time (using a standard desktop computer).
In \figref{fig:Fig6} we compare the time-dependent behaviour of the exact full model and RM1, plotting the mean and standard deviation of $Y(t)$, $\mu_Y$ and $\sigma_Y$ respectively. We initialise the system with $X(0) \sim \rho(x)$ and  $Y(0) = 0$. While the mean obtained from the RM1 approximation is indistinguishable from that of the full system, the standard deviation is slightly underestimated by RM1. We note that the error in $\sigma_Y$ in \figref{fig:Fig6}(b) is in fact quite small relative to the mean $\mu_Y$ (0.65\%), and this is why this difference is only just recognisable in \figref{fig:Fig5}.

\subsection{Regime 2: $\epsilon\ll \tau_s \sim \tau_y$} \label{sec:rm2}

When the timescale for the evolution of species $Y$ is of the same
order as that of  the switches in $X(t)$, it is important to keep the
bistable nature of the system in the reduced model, since $Y$ responds differently depending on which well $X$ is in.
\figref{fig:Fig4}(b) shows an example of the system in this regime. As in the one-dimensional example described in \secref{sec:intro}, using a QSA the bistability can be kept in by introducing a discrete two state stochastic process $S(t)$ governing the jumps of $X(t)$: $S(t) = S_-$ when $X(t)$ is in the left well $\Omega_- =  [0  , x_*)$ and $S(t) = S_+$ when $X(t)$ is in the right well $\Omega_+ =  (x_*, \infty)$. Within each well $X$ quickly reaches a quasi-equilibrium, and therefore we can use a modified stochastic averaging conditional on the value of $S$. Then the reduced stochastic system is  
\begin{subequations}
\label{reducedmodel2}
\begin{equation}
\label{reactiony_reg2}
\ce{$S_-$  <=>[k_-][k_+] $S_+$}, \qquad
\ce{$\emptyset$  <=>[{\overline k_5}(S)/\tau_y][k_6/\tau_y] $Y$}
\end{equation}
where 
the effective production rate takes one of two values depending on
whether $X$ is in the left or right well, 
\begin{equation}
 {\overline k_5}(S_\pm) = \int_{\Omega_\pm } k_5 (x) \rho_\pm(x) \, \ud x;
\end{equation}
here $\rho_{\pm}$ is the normalised stationary density of $X$
conditional on being in the left (minus) or right (plus) well. This definition corresponds to imposing a reflecting boundary condition on $x_*$ in \eqref{fokker-planck_x}.\footnote{This is in contrast to the Parallel Replica Algorithm,\cite{LeBris:2012et} which defines quasi-stationary distributions using absorbing boundary conditions.}
A formal derivation of this model is given in \ref{sec:regime2}.
We denote this reduced system by \emph{reduced model 2} or RM2. The rate constants $k_\pm$ of the process for $S(t)$ are the inverse of the mean switching times $T_\pm$ for $X$ to jump from one well to the other [the switching timescale $\tau_s$ is such that $T_\pm \sim O (\tau_s)$]. Determining these rates accurately is one of the main challenges of Regime 2; we will return to this issue in section \ref{sec:mst}. For the chemical $X(t)$, we will find that 
\begin{equation}
\label{mst_chem1}
k_- = 0.0931, \qquad k_+ = 0.0835. 
\end{equation}
\end{subequations}

\begin{figure}[htb]
\begin{center}
\includegraphics[width=.9\columnwidth]{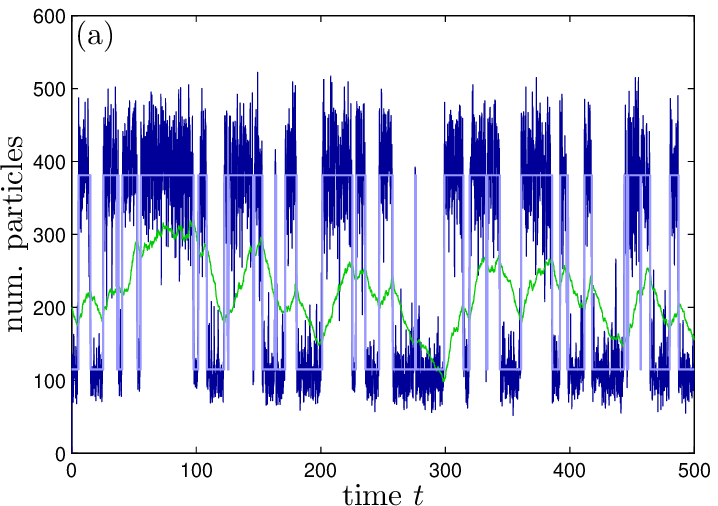} \\ \includegraphics[width=.9\columnwidth]{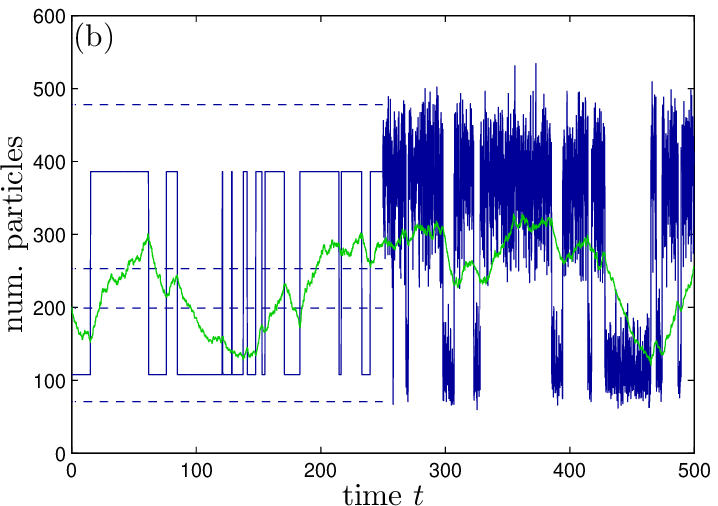} 
\caption{Comparison of one single realisation in Regime 2 of (a) the exact full system \eqref{reactions} and (b) the approximated reduced model 2 \eqref{reducedmodel2} using the NRM algorithm. We use $\tau_y = 25$, $X(0)= 0$ [\ie $S(0) = S_-$ in (b)], $Y(0) = 200$, and the parameter values \eqref{values}. In (a), the simulated quantities $X(t)$ and $Y(t)$ are shown in dark blue and green, respectively. The lighter blue line shows reduced two-valued curve $\overline X_\pm$ corresponding to the bistable switch [$\overline X_\pm$ if $X(t) \in \Omega_\pm$, see \eqref{twovalued}]. In (b), the simulated variable $Y(t)$ is shown in green, and the approximate representation of $X(t)$ from the simulated $S(t)$ in shown as follows: in the first half, $\overline X_\pm$ (according to $S(t) = S_\pm$) and the 95\% confidence intervals of $\rho_\pm(x)$ are plot as solid and dash blue lines, respectively. In the second half, the solid blue line shows samples taken every $\Delta t = 0.1$ from $\rho_\pm(x)$ as per $S(t) = S_\pm$.}
\label{fig:Fig7}
\end{center}
\end{figure}

A comparison of a sample trajectory of the system obtained from a NRM simulation of the full system and of the reduced system RM2 is shown in \figref{fig:Fig7}. We
choose the timescale of $Y(t)$ such that $\tau_y \sim \tau_s$ and the
system is in Regime 2. In \figref{fig:Fig7}(a), the $X(t)$ and $Y(t)$ populations simulated from the full model \eqref{reactions} are shown in dark blue and green solid lines, respectively. The two-valued light blue curve represents the bistable switching of $X(t)$ and takes as values the conditional average values 
\begin{equation}
\label{twovalued}
\overline X_\pm = \int_{\Omega_\pm} x \rho_\pm(x) \, \ud x,
\end{equation}
according to when $X(t) \in \Omega_\pm$. \figref{fig:Fig7}(b) shows one trajectory of the reduced model \eqref{reducedmodel2}. In RM2, $X(t)$ is not explicitly simulated, but we can still illustrate an approximate trajectory using the switch variable $S(t)$: in the first half of \figref{fig:Fig7}(b), we plot $\overline X_\pm$ as specified by $S(t) = S_\pm$ as well as the corresponding  95\% confidence interval of $\rho_\pm(x)$ (dash blue lines). In the second half, to allow for an easier comparison with \figref{fig:Fig7}(a),  
we plot instead samples $X$ from $\rho_\pm(x)$ depending on the value of $S(t)$ (taken every $\Delta t= 0.1$). Using this procedure, we can make a run of the RM2 look very similar to a full model run as in \figref{fig:Fig7}(a). 

\begin{figure}[hbt]
\begin{center}
\includegraphics[width=\columnwidth]{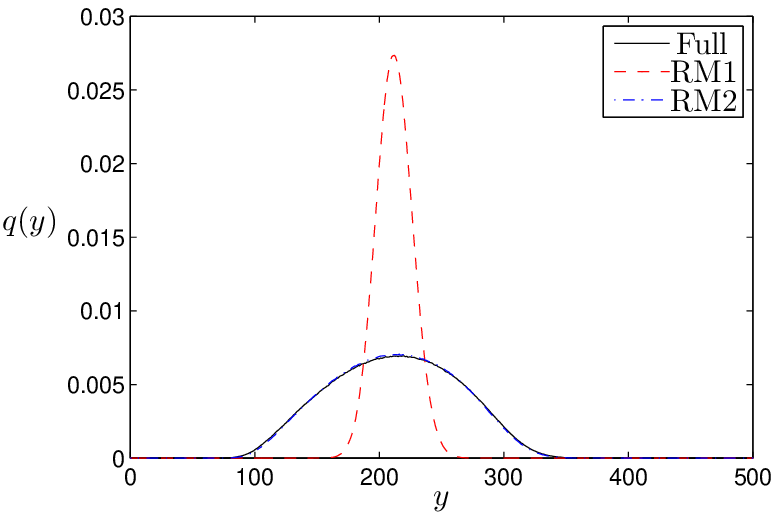} 
\caption{Comparison of the marginal stationary density $q(y)$ in regime 2: full system \eqref{reactions} (solid black line) and approximate reduced models 1 \eqref{reactiony_reg1} (dash red line) and 2 \eqref{reducedmodel2} (dot-dash blue line). All three histograms are computed by running the NRM algorithm up to $t = 10^7$ taking recordings every $\Delta t = 0.5$ ($2\times 10^7$ samples). We use $\tau_y = 25$  and the parameter values \eqref{values}.}
\label{fig:Fig8}
\end{center}
\end{figure}
\begin{figure}[bht]
\begin{center}
\includegraphics[width=.9\columnwidth]{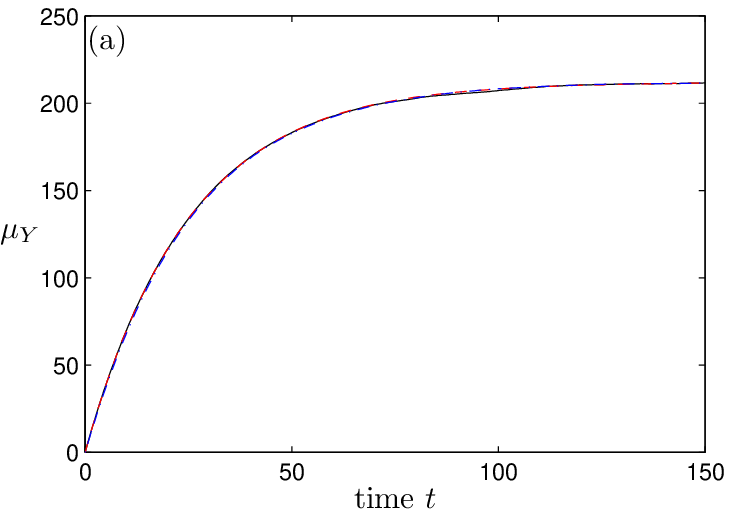} \\ 
\includegraphics[width=.9\columnwidth]{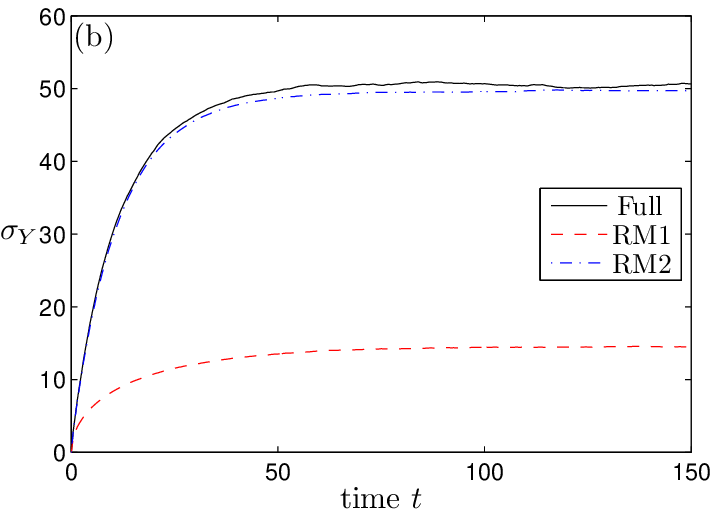}
\caption{Comparison of the time-dependent (a) mean $\mu_Y(t)$ and  (b) standard deviation $\sigma_Y(t)$: full system \eqref{reactions} (solid black line) and approximate reduced models 1 \eqref{reactiony_reg1} (dash red line) and 2 \eqref{reducedmodel2} (dot-dash blue line). Black, red, and blue curves are computed as the average over $1.1\times 10^4$,  $10^5$, and $10^5$ realisations respectively with initial conditions $Y(0) = 0$ and $X(0) \sim \rho(x)$. We use $\tau_y = 25$ and the parameter values in \eqref{values}.}
\label{fig:Fig9}
\end{center}
\end{figure}

While \figref{fig:Fig7} indicates that the qualitative
behaviour of the
reduced model \eqref{reducedmodel2} is 
similar to that of the full system, a more quantitative
comparison is appropriate.
In \figref{fig:Fig8} we compare the stationary distributions of $Y$ (histograms) obtained from a long-time NRM simulation of the full model \eqref{reactions} and the reduced model 2 \eqref{reducedmodel2}. 
To illustrate the need for a new model reduction in Regime 2, we also plot the histogram obtained by a standard stochastic averaging using RM1 \eqref{reactiony_reg1}. The RM2 solution is in excellent agreement with the full system, while the RM1 approximation considerably underestimates the variance of the distribution. 
In \figref{fig:Fig9} we show a comparison of the mean and standard deviation of $Y(t)$ for the three models. We initialise the system with $Y(0) = 0$ and $X(0) \sim \rho(x)$. Note that the latter is not required for RM1 since $X(t)$ has been eliminated from that model; in contrast, the initialisation for RM2 is $S(0) = S_\pm$ depending on the well in which the sample from $\rho(x)$ is in [namely, $S(0) = S_-$ with probability $\int_{\Omega_-} \rho(x) \ud x$, and $S_+$ otherwise]. 
We see that while the RM1 \eqref{reactiony_reg1} captures the behaviour of the mean of $Y$ fairly well, it significantly underestimates the variance of $Y$.

\subsection{Regime 3: $\epsilon\sim \tau_y \ll \tau_s$} \label{sec:rm3}
\begin{figure}[thb]
\begin{center}
\includegraphics[height=.3\textwidth]{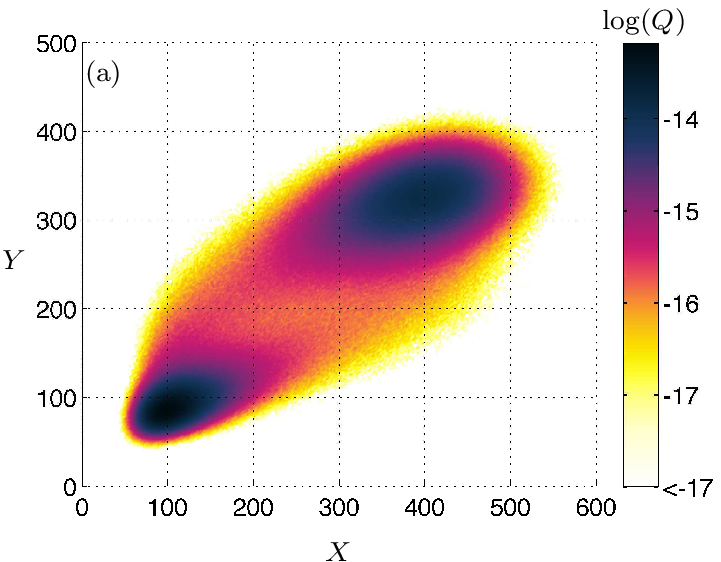} \\
\includegraphics[height=.3\textwidth]{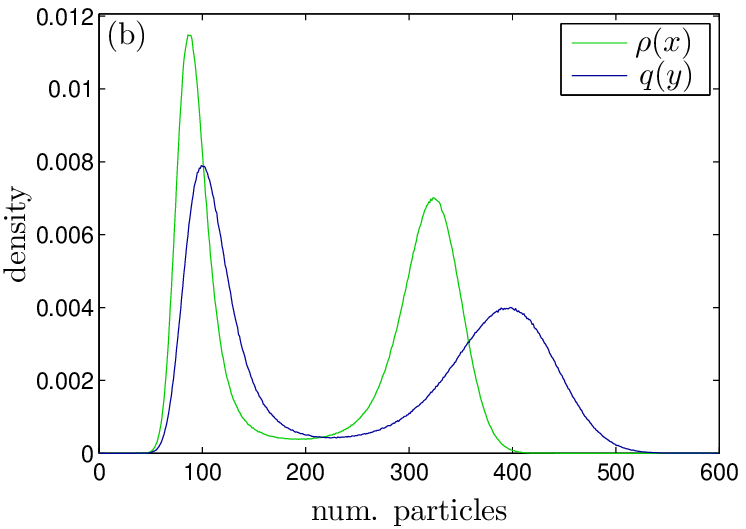} 
\caption{(a) Logarithm of the joint stationary density $Q(x,y)$ and (b) marginal stationary densities $\rho(x)$ and $q(y)$ measured from a long simulation (up to $t=10^6$ taking recordings every $\Delta t = 0.1$) of the full system \eqref{reactions} (blue and green histograms, respectively). We use $\tau_y = 0.25$ and  the parameters values \eqref{values}.} 
\label{fig:Fig10}
\end{center}
\end{figure}

Now consider a scenario in which $\tau_y \ll \tau_s$, as illustrated
in \figref{fig:Fig4}(c). 
Species $Y$ is now fast by comparison to the switches of $X$ between
wells. Thus on the slowest timescale of $\tau_s$ both $X$ and $Y$ have
time to reach a quasi-equilibrium, conditional on $X$ being in a given
well. In this case, when we eliminate the fast variables we are  left
only with the binary switch $S(t)$. Hence the reduced system in Regime 3 on the $\tau_s$
timescale is simply  
\begin{equation}
\label{reactiony_reg3}
\ce{$S_-$  <=>[k_-][k_+] $S_+$},
\end{equation}
where $k_\pm$ are given in \eqref{mst_chem1}; see \ref{sec:regime3} for more details. We call this the \emph{reduced model 3} (RM3). When $S(t) = S_-$ (resp. $S_+$), the system has a quasi-stationary density $Q_-(x, y)$ [resp. $Q_+(x, y)$], which is the stationary density conditional on $X$ being in the left (resp. right) well. To clarify this, in \figref{fig:Fig10} we plot the joint stationary density $Q(x,y)$ and the two marginal stationary densities $\rho(x)$ and $q(y)$ obtained by the long-time exact simulation of the full system \eqref{reactions} with $\tau_y$ such that the system is in Regime 3. This confirms that $Y(t)$ is also bimodal in this regime [even though the switches between its two modes are still controlled by $X(t)$]. To obtain the quasi-stationary densities $Q_\pm(x,y)$ [and the corresponding marginals $\rho_\pm(x)$ and $q_\pm(y)$], we follow a similar proceeding using a long-time simulation of \eqref{reactions} but with a reflecting boundary condition at $x_*$.

\begin{figure}[hbt]
\begin{center}
\includegraphics[width=\columnwidth]{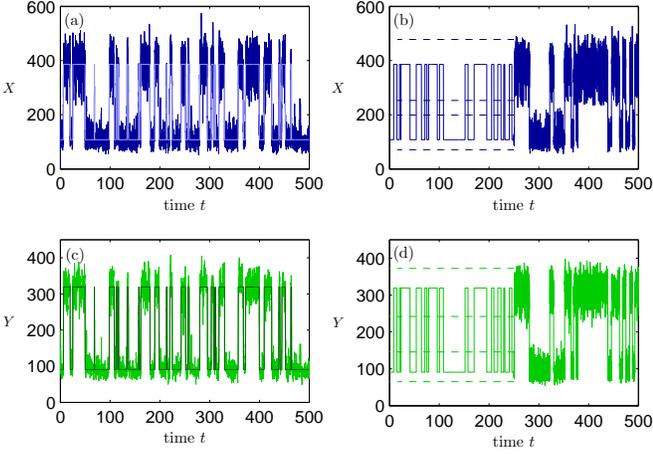} 
\caption{Comparison of one realisation in Regime 3 of (a,c) the exact full system \eqref{reactions} and (b,d) the approximated reduced model 3 \eqref{reactiony_reg3} using the NRM algorithm. We use $\tau_y = 0.25$, $X(0)= 0$ [\ie $S(0) = S_-$ in (b,d)], $Y(0) = 200$, and the parameter values \eqref{values}. In (a,c), the simulated quantities $X(t)$ and $Y(t)$ are shown in dark blue and light green, respectively. The two-valued curves in (a,c) are computed from \eqref{reg3_filter}. In (b,c), the approximate representation of $(X(t), Y(t))$ from the simulated $S(t)$ in shown as follows: in the first half, \eqref{reg3_filter} and the 95\% confidence intervals of $\rho_\pm(x)$ and $q_\pm(y)$ are plot as solid and dash lines, respectively. In the second half of (b) and (d), curves show samples of $\rho_\pm(x)$ and $q_\pm(y)$, respectively, taken every $\Delta t = 0.1$ from as per $S(t) = S_\pm$.}
\label{fig:Fig11}
\end{center}
\end{figure}

In \figref{fig:Fig11} we compare a sample trajectory of the full set of reactions \eqref{reactions} with sample trajectory of the reduced system \eqref{reactiony_reg3}. We choose $\tau_y = 0.25$ so that the system is in Regime 3.  Figures \ref{fig:Fig11}(a) and (c) show the evolution of $X(t)$ and $Y(t)$ respectively from \eqref{reactions}, together with the two-valued filtered curves to represent the switches in $X(t)$ given by  
\begin{equation}
\label{reg3_filter}
(X(t), Y(t) ) = \!
\begin{cases}
(\overline X_- , \overline Y_-),  & \text{if} \ \, S(t) = S_-, \ (X(t) < x_*),\\
(\overline X_+ , \overline Y_+), & \text{if}\ \, S(t) = S_+, \ (X(t) > x_*),
\end{cases}
\end{equation}
where 
\begin{equation}
\label{conditional_mean}
\overline X_\pm = \int_{\Omega_\pm} x Q_\pm(x, y) \, \ud x \ud y,
\end{equation}
and similarly for $\overline Y_\pm$.  \figref{fig:Fig11}(b) and (d) show one trajectory of the RM3 \eqref{reactiony_reg3}, where the output $S(t)$ is mapped to $X(t)$ and $Y(t)$ units using the quasi-stationary densities $\rho_\pm(x)$ and $q_\pm(y)$ [similarly to what we did in \figref{fig:Fig7}(b)]. In the first half of Figs. \ref{fig:Fig11}(b,d) we plot $(X(t), Y(t) )$ \eqref{reg3_filter} (solid lines) and the 95\% confidence intervals (dashed lines) of $\rho_\pm(x)$ and $q_\pm(y)$. In the second half, instead of the average quantities we plot sampled values $(X,Y)$ from $Q_\pm(x,y)$ [depending on the value of $S(t)$]. This leads to qualitatively very similar output as with the full model in the left column of \figref{fig:Fig11}. 

\begin{figure}[htb!]
\begin{center} 
\includegraphics[width=\columnwidth]{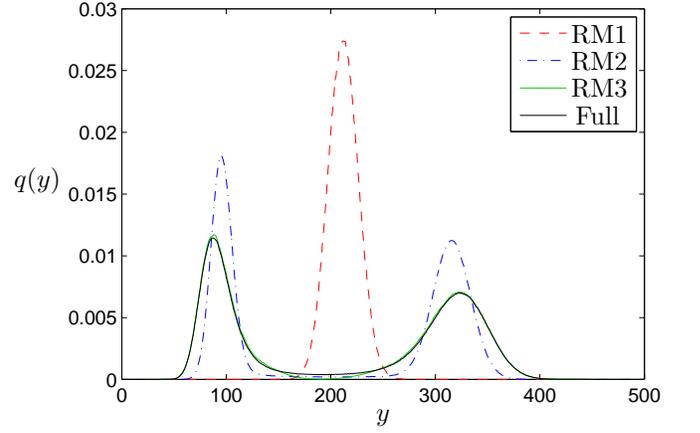} 
\caption{Comparison of the marginal stationary density $q(y)$ in regime 3: full system \eqref{reactions} (solid black line) and approximate reduced models 1 \eqref{reactiony_reg1} (dash red line), 2 \eqref{reducedmodel2} (dot-dash blue line), and 3 \eqref{reactiony_reg3} (solid green line). All four histograms are computed by running the NRM algorithm up to $t = 10^6$ taking recordings every $\Delta t = 0.1$ ($10^7$ samples). We use $\tau_y = 0.25$  and the parameter values \eqref{values}.}
\label{fig:Fig12}
\end{center}
\end{figure}

In the next two figures we show a quantitative comparison in Regime 3 between the full system and all three reduced models. In \figref{fig:Fig12} we compare the stationary distribution of $Y$ obtained by long-time simulations of the four models. We note an excellent agreement between the histogram from RM3 (solid green line) and that from the full model (solid black line). But this is not surprising since to get $q(y)$ with the RM3 we use the quasi-steady densities $q_\pm(y)  = \int_{\Omega_\pm} Q_\pm(x, y) \, \ud x$ (obtained in turn from a full model simulation with a reflecting condition at $x_*$), and the only output from RM3 is the proportion of time spend in the left/right wells. The histogram from RM2 (dot-dash blue line) captures the bimodality of $Y$ in this regime, but it gets the conditional means and variances substantially wrong. This is because $X$ and $Y$ are varying in the same timescale whereas RM2 supposes that $X$ is much faster than $Y$. If the timescales in Regime 3 satisfied $\epsilon \ll \tau_y \ll \tau_s$, we would expect RM2 to give a good approximation to the full model dynamics. 
\begin{figure}[hbt]
\begin{center}
\includegraphics[width=.9\columnwidth]{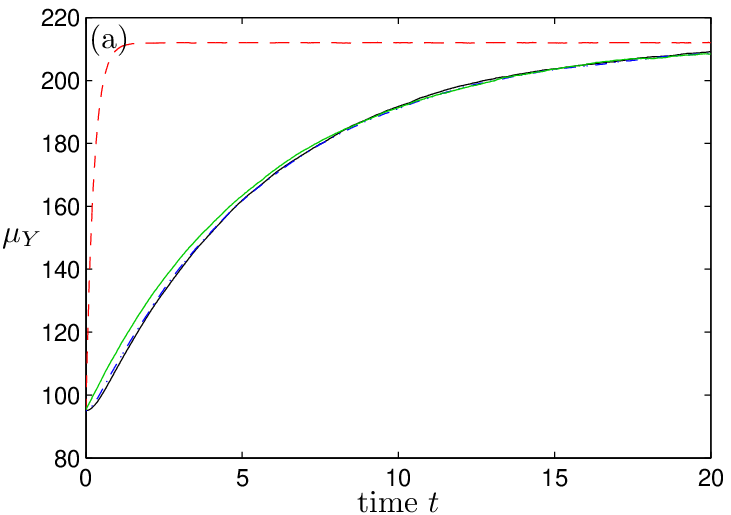} \\
\includegraphics[width=.9\columnwidth]{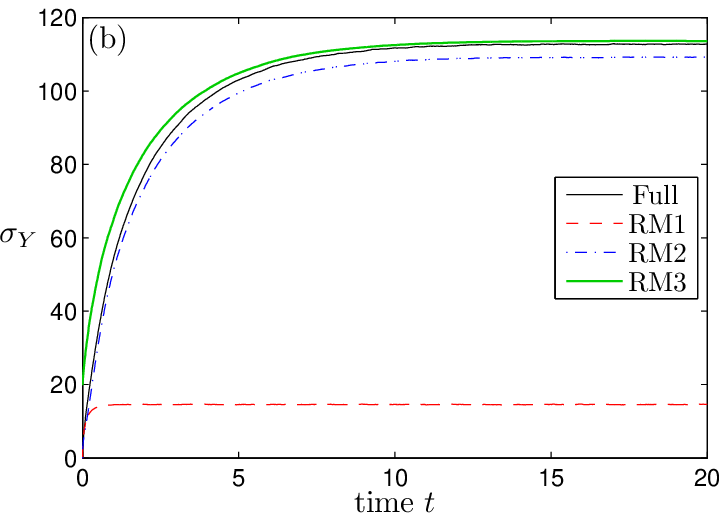}
\caption{Comparison of the time-dependent (a) mean $\mu_Y(t)$ and  (b) standard deviation $\sigma_Y(t)$: full system \eqref{reactions} (solid black line) and approximate reduced models RM1 \eqref{reactiony_reg1} (dash red line), RM2 \eqref{reducedmodel2} (dot-dash blue line), and RM3 \eqref{reactiony_reg3} (solid green line). All curves are computed as the average over $10^5$ realisations with initial conditions $X(0) \sim \rho_-(x)$ and $Y(0) = \overline Y_- \approx 95$. We use $\tau_y = 25$ and the parameter values in \eqref{values}. The curves for RM3 are computed using \eqref{musig-reg3}.}
\label{fig:Fig13}
\end{center}
\end{figure}

\figref{fig:Fig13} shows the mean $\mu_Y(t)$ and the standard deviation $\sigma_Y(t)$ of $Y$ obtained from $10^5$ realisations of the four models. We initialise the system in the left well ($S(0) = S_-$), using $X(0) \sim \rho_-(x)$ and $Y(0) \sim q_-(y)$ for the models that require explicit $X$ or $Y$ initialisation. The values of $\mu_Y(t)$ and $\sigma_Y(t)$ corresponding to RM3 can be calculated from $S(t)$. A simple calculation shows that they are given by
\begin{equation}
\label{musig-reg3}
\begin{aligned}
\mu_Y(t) &= \overline Y_- + \left( \overline Y_+ - \overline Y_- \right) s_+(t),\\
\sigma^2_Y(t) &= \sigma^2_{Y_-} + \left[ \sigma^2_{Y_+} - \sigma^2_{Y_-} \right] s_+(t) \\
& \phantom{=\,} + \left [ \overline Y_+ - \overline Y_- \right]^2 s_+(t) [1-s_+(t) ],  
\end{aligned}
\end{equation}
where $s_+(t)$ is the estimated probability that $S(t) = S_+$ and $\sigma^2_{Y_\pm} = \int (y- \overline Y_\pm)^2 q_\pm(y) \ud y$.
The mean $\mu_Y$ of both RM2 and RM3 captures well the mean of the full model, while RM1 misses the transient and jumps quickly to its stationary value (since it is taking $X$ to be equilibrated instantly at $t=0$). Similarly, the global standard deviation $\sigma_Y$ of the full model is well approximated by RM2 and RM3, while RM1 underestimates (this is to be expected as we have eliminated the noise coming from the switches of $X$).

\subsection{Estimation of mean-switching times}
\label{sec:mst}

In reducing the fast variable $X$ to a two-state Markov process in the reduced models 2 and 3 (corresponding to taking the limit  $\delta \rightarrow 0$ as we will see in Appendix \ref{sec:regime2}), the key pieces of information we need to extract are the switching rates $k_+$ and $k_-$. These are the inverses of the mean transition (or escape) times $T_+$ and $T_-$. In this section we show how to accurately obtain these rates for the chemical system. 
 
The estimation of the mean escape times for metastable processes is a classical problem that has received much attention in the literature.\cite{Risken:1996vl} As $\delta \to 0$ the escape times become exponentially small in $\delta$  and may be estimated by a variety of techniques in exponential asymptotics.\cite{Hinch:2005eg, Ward:1998ve}
In some cases the process for $X$ may be too complicated to estimate the mean escape times analytically, and a numerical estimate must be used.\cite{LeBris:2012et} However, when $\epsilon$ and $\delta$ are small but non-zero even the definition of escape becomes an issue: at what stage has the process $X$ reached the other well?

For our simple bistable $X(t)$ process \eqref{sde_x}, the mean time to reach any given point $x_1$, given we start at $X(0) = x_0$ is given exactly by\cite{Erban:2009ew} 
\begin{equation}
\label{met_exact}
\begin{aligned}
T_-(x_0, x_1) & = \int_{x_0} ^{x_1} \int_{0}^{z} \frac{\rho(s)}{d(z) \rho(z)} \ud s \ud z, \quad \ \text{with} \quad x_0 < x_1, \\
T_+(x_0, x_1) &= \int_{x_1} ^{x_0} \int_{z}^{\infty} \frac{\rho(s)}{d(z) \rho(z)} \ud s \ud z, \quad \text{with} \quad x_0 > x_1.
\end{aligned}
\end{equation}
The question is, what values do we choose for $x_0$ and $x_1$?

For $x_0$ we could choose the local maximum of $\rho$ in the left-hand
well (which we denote by $x_-$), or we could sample from the
stationary distribution  conditional on being in the left-hand
well (which we denote by $\rho_-$). Since in the limit $\epsilon\ll
\tau_s$ equilibration within a well is rapid by comparison to
transitions between wells, we could in principle start with any value
of $x_0$ in 
the left-hand well and we would obtain the same transition time to
leading order.

For $x_1$ we could use
$x_*$, the minimum of $\rho$, which satisfies
\[ d'(x_*) - v(x_*) = 0.\]
We then need to double the mean first passage time to find the mean escape
time, since a particle at $x_*$ 
will return to the well it 
came from with probability one half.
Alternatively, since equilibration within a well is rapid by
comparison to transitions between wells, we may choose any $x_1$ which
is sufficiently far from $x_*$ so that immediate return to the
left-hand well is unlikely. For example, we could choose the mean
$(x_*+x_+)/2$, where $x_+$ is the local maximum of $\rho$ in the
right-hand well.\cite{Erban:2009ew}

\begin{table*}
\caption{\label{table1} Mean exit times $T_-$ and $T_+$ to leave the left and right
  wells, respectively, as derived from numerical  simulations and theory. We compare the results
  obtained using different start and end points.  } 
\begin{ruledtabular}
\begin{tabular}{c c c c | c c c c}
  \multicolumn{4}{c}{$T_-$ (left $\to$ right)} &  \multicolumn{4}{c}{$T_+$ (right $\to$ left)} \\
  \hline 
 $x_0$& $x_1$& Eq. \eqref{met_exact} & SSA &  $x_0$& $x_1$& Eq. \eqref{met_exact} & SSA\\ \hline
$\rho_-(x_0)$ & $(x_* + x_+)/2$ & 11.1190  & 11.1275 & $\rho_+(x_0)$ & $(x_- +  x_*)/2$ & 11.5366  & 11.5573   \\
$x_-$ & $(x_* + x_+)/2$ & 11.4443 & 11.4610 & $x_+ $ & $(x_- + x_*)/2$ & 11.8714 & 11.9482  \\ 
0 & $(x_* + x_+)/2$ & 11.6121  & 11.3253   & 1000\footnotemark[1] & $(x_- + x_*)/2$ & 11.9853   & 12.1354  \\
 0 & $x_*$\footnotemark[2] & 13.7263 & 13.7463  & 1000\footnotemark[1]  & $x_*$\footnotemark[2] & 11.2982  & 11.5974   \\ 
\end{tabular} 
\end{ruledtabular}
\footnotetext[1]{The
  starting point for the right to left transitions, which in theory should be $x = \infty$, is placed at $x = 1000$, where $\rho_+(1000) \sim 10^{-59}$. The relative error in the theoretical values using $\infty$ or 1000 are $O(10^{-4})$.}
\footnotetext[2]{The values
  $T_\pm$ using $x_1 = x_*$  are multiplied by two (both in
  the theory and SSA) since there is, roughly speaking, a 50\% chance
  to fall either side once $X(t)$ reaches the unstable point $x_*$.}
\end{table*}

 In Table \ref{table1}  we present values of  $T_\pm$ [using \eqref{met_exact}] and the corresponding results obtained by the SSA (with relative standard error of less than $1\%$, achieved with roughly $10^4$ exits) for several choices of  $x_0$ and $x_1$. We see that there is a small but significant variation in the switching times despite the fact that $\epsilon$ is quite small. To decide which of these times to use we consider here what properties
we require of the reduced system \eqref{reactiony_reg2}.
 If we denote $s_\pm(t)$ as the probability that $S(t) = S_\pm$ then 
$s_+(t)$ satisfies
\begin{equation}
\frac{\ud s_+}{\ud t} = k_- - (k_- + k_+) s_+,
\end{equation}
where we have used the relation $s_- + s_+ =1$ to eliminate $s_-$.
If we initialise $X$ in the left-hand well [by sampling from
$\rho_-(x)$] then $s_+(0) = 0$, giving 
\begin{equation}
\label{eq_sright}
s_+(t) = \frac{k_-}{k_- + k_+} \left ( 1 - e^{-(k_- + k_+) t} \right).
\end{equation}
The stationary value $s_+ = \frac{k_-}{k_- + k_+}= 1-\theta$,  represents the
proportion of time that $X$ spends in the right-hand well on average.
For $S$ to be a good approximation to $X$ this should be equal to the
integral of $\rho(x)$ over the right-hand well.
This gives us one relationship between $k_+$ and $k_-$, which
determines the ratio between the mean switching times $T_\pm$. It can be
shown that choosing $k_+$ and $k_-$ which satisfy this constraint will
ensure that the reduced process accurately captures the mean behaviour
of $Y$. 

To accurately capture the variance in $Y$ we need to capture the rate
of approach to this stationary solution accurately, i.e. we need to
determine the time constant  $\psi = 1/(k_- + k_+)$.
We can avoid the difficulty of determining when a switch in $X$ has occurred by
considering the time dependent mean of both $X$ and the reduced
process $S$. If we can match the decay rates of these means, then we
will have determined the rate constant $\psi$ accurately.
The mean of $X$ under the reduced model is given by
\begin{equation}
\label{estimate_x}
\overline X(t) = \theta \overline X_- +(1- \theta) \overline X_+ -
(1-\theta) (\overline X_+ - \overline X_-) e^{- t/\psi}, 
\end{equation}
where $\overline X_\pm = \int_{\Omega_\pm} x \rho_\pm (x) \, \ud x$
are the mean values of $X$ restricted to the right and left well
respectively. The equilibrium value is $\overline X_\infty = \theta \overline X_- +(1- \theta) \overline X_+$. 
By comparing \eqref{estimate_x} to  an ensemble of short-time 
stochastic simulations of the full process $X(t)$ we are able to get a
good estimate $\psi$, 
which is then enough to determine $k_-$ and $k_+$.

\begin{figure}[hbt]
\begin{center}
\includegraphics[width=.4\textwidth]{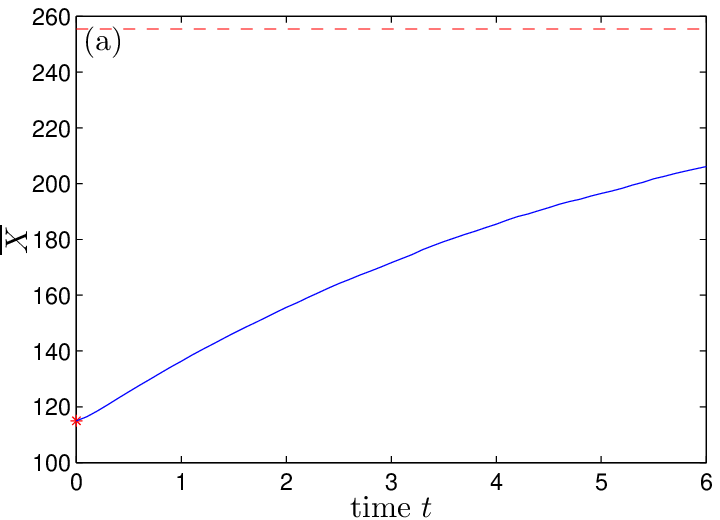} \\ 
\includegraphics[width=.4\textwidth]{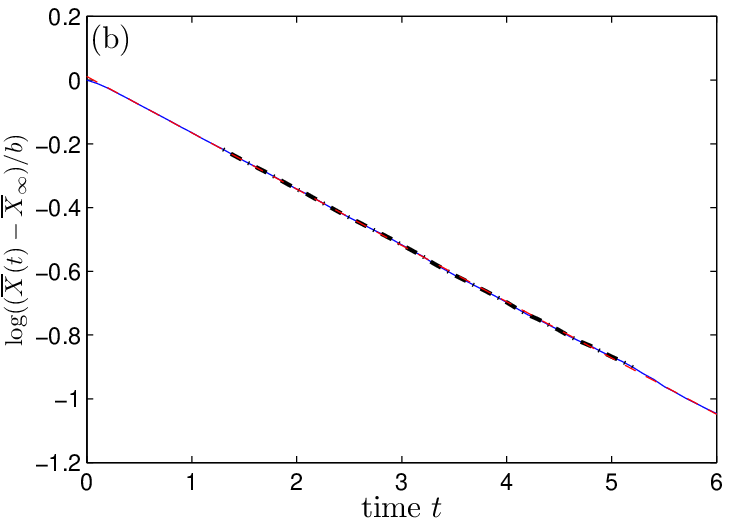} 
\caption{Estimation of $\psi$ from \eqref{estimate_x}. (a) Time-evolution of the mean $\overline X (t)$ with $\overline X(0) = \overline X_-$ (solid blue line). The stationary value $\overline X_\infty$ is shown as a dash red line. The curve has been obtained from $10^5$ SSA realisations of the full model \eqref{reactions}. (b) Transformation $\log ((\overline X -  \overline X_\infty)/b)$ with $b =  (\theta-1) (\overline X_+ - \overline X_-)$ using $\overline X(t)$ in (a). The parameter $\psi$ is estimated from the slope of this curve, using \eqref{estimate_x}. The black dashed line shows the section we use for the linear fit (shown in as a dot-dashed red line).}
\label{fig:Fig14}
\end{center}
\end{figure}
In  Figure \ref{fig:Fig14}(a) we show the computed time evolution of $\overline{X}$, the mean of the full $X$ process (obtained as the average of $10^5$ runs).
If from this we subtract the large time behaviour $\overline{X}_\infty$ and then take a logarithm we should obtain a straight line with gradient $-1/\psi$, as shown in \figref{fig:Fig14}(b). We fit a straight line
to the part of the curve in which $\overline{X}-\overline{X}_\infty$   lies between
80\% and 40\% of its initial value, in order to avoid any initial
transients.
The fit is good, and leads to the following values for $\theta,
\psi$ and $T_\pm$:  
\begin{equation}
\begin{aligned}
\theta &= 0.4729, & \quad \psi &= 5.6602, \\
 T_-  &= 10.7379,& T_+  &= 11.9697. 
\end{aligned}
\end{equation}
We see that these are in the same range as those in Table~\ref{table1}. These are the values of $T_\pm$ that have been used in all simulations for RM2 and RM3 presented in Subsections \ref{sec:rm2} and \ref{sec:rm3}.

\section{Calculating extinction times in a predator--prey system}
\label{sec:modelproblem2}

In this section we apply the model reduction methods developed above to an ecological model. In particular, we  consider the probability of extinction of a population of predators when the prey undergoes  a metastable stochastic process with bistability.
This case study allows us to summarise the application of the method and, in particular, the two ingredients that one must extract from the original slow--fast system, namely the quasi-stationary densities and the mean switching times. 
Most importantly, this model invites us to push the method further by examining its performance with a system with absorbing states. We are thus interested in capturing the evolution of the system for low numbers. As a result, rather than using the continuous Fokker--Planck (FP) approximation for both species as we did for the chemical system, in this section we use the discrete description based on the backward master equation for the dynamics of the slow variable whose extinction we want to study. 

\begin{table*}[ht]
\caption{\label{table2} Propensity functions.}
\begin{tabularx}{.6\textwidth}{llll}
\hline\hline
\multicolumn{2}{c}{Prey} & \multicolumn{2}{c}{Predator} \\
\hline
Transition & Propensity $\alpha_i$&Transition & Propensity $\alpha_i$\\
\hline
$X \to X+1$ & $\alpha_1 (x) = \lambda x - \lambda \frac{x^2}{\kappa}$ &$Y \to Y+1$ & $\alpha_3(y) =\tilde \lambda y$ \\ 
$X \to X-1$ & $\alpha_2(x) =\mu x + \frac{\beta  z x^2}{1 + \beta h x^2}$ \hspace{1cm} &$Y \to Y-1$ & $\alpha_4 (x,y) = \tilde \mu y +  \frac{\tilde \beta y^2}{K(x)}$ \\ 
 \hline\hline
 \end{tabularx}
\end{table*}%

\begin{figure*}[thb]
\begin{center}  
\includegraphics[width=.32\textwidth]{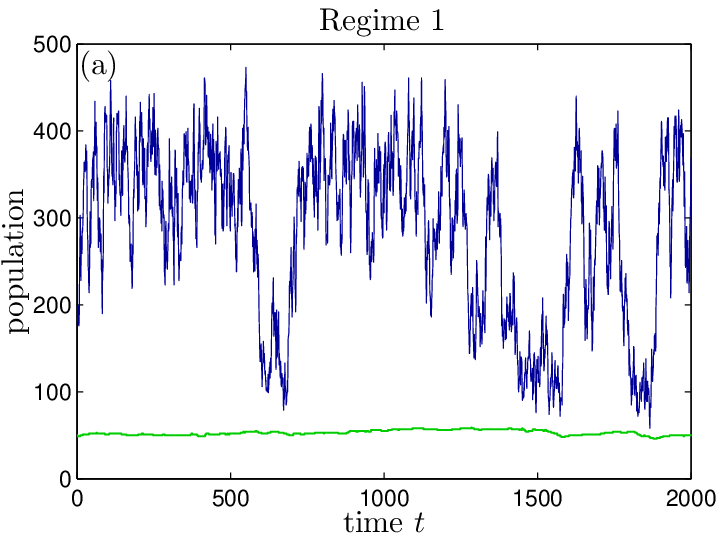}  
\includegraphics[width=.32\textwidth]{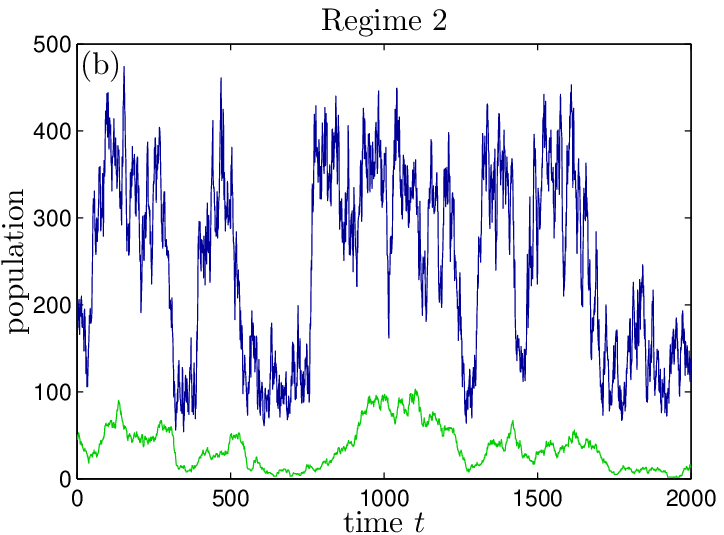} 
\includegraphics[width=.32\textwidth]{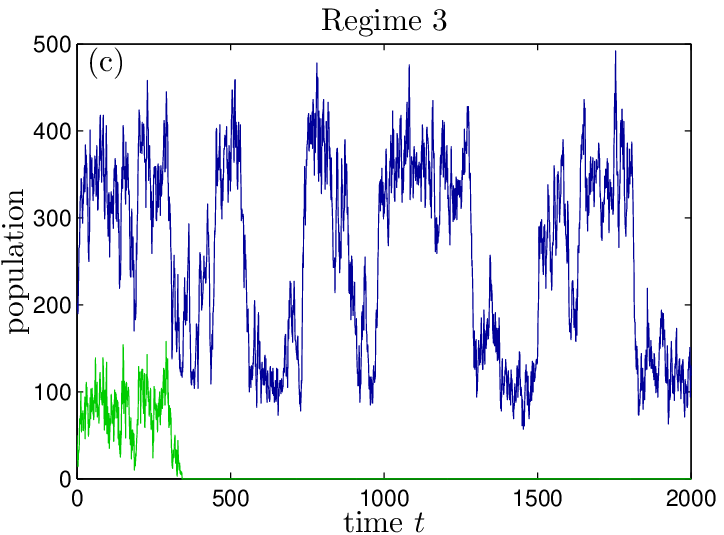} 
\caption{Evolution of the prey $X(t)$ (blue lines) and the predator $Y(t)$ (green lines) from a single realisation of \eqref{reactions_pp} obtained using the NRM algorithm. We use the parameter values \eqref{parameters_pp}, $X(0) =200$, $Y(0) = 50$, and vary $\tau_y$ to change between regimes.  (a) Regime 1: $\tau_y = 10^4$. (b) Regime 2: $\tau_y = 100$. (c) Regime 3: $  \tau_y = 1$. }
\label{fig:Fig15}
\end{center}
\end{figure*}

We briefly give some background to our choice of model for this case study. One of the most fundamental questions in population biology concerns the persistence of species and populations, or conversely their risk of extinction.\cite{Ovaskainen:2010wm} Stochastic population models have become a common tool to investigate how the mean time to extinction depends on properties of the ecosystem. 
However, performing detailed mathematical investigations to understand how extinction risk depends on properties of the ecosystem is limited by the availability of tractable yet relevant models. A number of simple stochastic models of population dynamics have been used to study the effects of demographic processes on the mean time to extinction.\cite{Ovaskainen:2010wm} A small subset of these have looked at the situation in which a population exhibits bistability (\eg Refs.~\onlinecite{Brassil:2001eg,Dennis:2002iy,Schreiber:2003hf}). These are predominantly cases in which there simultaneously exists a positive and a zero abundance attractor (locally stable steady states in the deterministic model). However, Palamara \etal\cite{Palamara:2012uoa} recently studied the mean time to extinction in simple predator--prey stochastic systems and reported a region of parameter space in which the prey population is bistable because there exists two alternative attractors with positive abundance, although they did not investigate that region for the mean time to extinction. We choose to use the model in Ref.~\onlinecite{Palamara:2012uoa} in the bistable regime as the basis for our case study because the population dynamics exhibited by the prey in this region are closely analogous to the bistable dynamics of $X$ in the chemical reaction case (\eg \figref{fig:Fig15}). To this we add a ``predator'' population whose abundance is influenced by the abundance of the prey, but does not itself influence the prey population (for example it survives on the direct by-products of the prey, such as dung beetles; while this is not strictly predation we will use the terms ``predation'' and ``predator'' here for clarity). In this case study we consider the probability of extinction of such a predator population when the prey undergoes a metastable stochastic process with bistability. 
We do not choose this example to represent any actual predator--prey system but rather to illustrate the advantages of using our model reduction techniques to enable the calculation of a property of domain-specific interest; in this case enabling efficient calculation of the probability of extinction of a population. 

We identify three regimes as in the previous chemical example, depending on the relative timescales for prey-population switches and predator dynamics, and use our reduction method to measure the extinction rate of the predator population without having to simulate the computationally costly full predator--prey system. 
We consider the following birth and death model for the prey\cite{Palamara:2012uoa},
\begin{subequations}
\label{reactions_pp}
\begin{equation}
\label{reactions_x2}
X \stackrel[]{k_1}{\longrightarrow} 2X, \hspace{2cm}  X  \stackrel[]{k_2}{\longrightarrow} \emptyset,
\end{equation}
where 
\[
k_1(x) = \lambda \left( 1- \frac{x}{\kappa} \right), \qquad k_2(x) = \mu + \frac{\beta x z}{1 + \beta h x^2}.
\]
Here the reproduction of $X$ corresponds to the Verhulst logistic model for population growth,\cite{Nasell:2001uc} where $\lambda$ is the intrinsic growth rate of the population and $\kappa$ is referred to as the carrying capacity of the environment. This abstraction is commonly used to represent the limiting-effects of population density on population growth, through mechanisms such as resource limitation.\cite{Turchin:2003tb}
The population death rate includes a constant death rate $\mu$ (predator-free death rate) and a predation-induced death rate. We use a Holling's Type III functional response, representing a situation in which predators consume multiple prey items and switch in their feeding preference to predating on species $X$ when it becomes particularly abundant.\cite{Turchin:2003tb} The term $\beta x$ is known as the attack rate, $h$ is the prey handling time and $z$ is the abundance of predators. 
Since we want this example to showcase the applicability of the method in a practical problem, here we do not introduce small parameters $\epsilon$ and $\delta$ as in the chemical system to make the characteristic timescales of the fast variable explicit.

We suppose that the predator pool for $x$ (parameterised by $z$) is composed of many species, of which we pick one, denoted by $y$. We assume that changes in the predator $y$ are negligible from the point of view of the prey, that is we can take $z$ to be constant. We take the predator population to evolve according to the  logistic model\cite{Grasman:1997vp} 
\begin{equation}
\label{reactions_y2}
Y \stackrel[]{k_3}{\longrightarrow} 2Y, \hspace{2cm} Y \stackrel[]{k_4}{\longrightarrow}  \emptyset,
\end{equation}
where
\[
k_3 = \tilde \lambda, \qquad k_4(x,y) = \tilde \mu +  \frac{\tilde \beta y}{K(x)},
\]
where $K(x)$ is a prey-dependent carrying capacity of the predator population. We will use $ K (x) = \tilde \kappa (1+ x^\theta)$ in what follows. 
The second term in $k_4$ accounts for competition for resources: the
death rate per individual increases with the predator population size
and decreases with increasing availability of prey. 
Note therefore that, from an ecological standpoint, $Y$ does not necessarily depend on $X$ to maintain a positive population size, but its abundance is influenced by the availability of $X$. See table \ref{table2} for a summary of the propensity functions for  each of the reactions. 
\end{subequations}

In \figref{fig:Fig15} we plot one run of the predator--prey
system \eqref{reactions_pp} for the following set of  parameters 
\begin{align}
\label{parameters_pp}
\begin{aligned}
\lambda &= 1.5, & \mu &=0.5,  & \beta &= 0.015/z, & \kappa &= 1000, \\
h &= 0.0055 z, & \tilde \lambda &= 5.5/\tau_y, &   \tilde \mu &= 3.5/\tau_y,  &  \tilde \beta &= 5.5/\tau_y, \\
 \tilde \kappa &= 0.002, &   \theta &= 2, & & & &
\end{aligned}
\end{align}
for three different timescales $\tau_y$ for the predator. As before, we vary $\tau_y$ relative to the prey switching timescale $\tau_s$ [which we find to be $\tau_s = O(10^2)$ in the next subsection, see \eqref{met_pp}] to move between regimes. 
\figref{fig:Fig15}(a) corresponds to $\tau_y = 10^4$ and is an example of a Regime 1 situation: the predator population hardly responds to fluctuations in the abundance of the prey. \figref{fig:Fig15}(b) shows a run with $\tau_y = 100$, which corresponds to Regime 2. Finally, in \figref{fig:Fig15}(c) the timescale of predator dynamics is set to be much shorter than the switching time of the prey by using $\tau_y = 1$, which fits with the definition of third regime.

We notice a new and interesting effect in this bistable system as we
vary the timescale of the population of interest and change the regime:
the predator population in Regime 3 [\figref{fig:Fig15}(c)] has
become extinct before the final simulation time $T_f = 2000$. This is only one run of the stochastic system, but if we repeat the same experiment many times a clear
pattern emerges: in Regime 3 the predator population reacts quickly to
a low level of prey population and its chances of becoming extinct
increase substantially. On the other hand, in Regime 1 the predator
population only sees an average of the prey population and not its low
and high levels and therefore the  carrying capacity always stays at
a level where time to extinction is long (since $\tilde \lambda  >
\tilde \mu$, giving a  basic reproduction ratio above one.
\cite{Nasell:2001uc}) In what follows we use the slow--fast
model reduction procedure to characterise the mean extinction time of
the predator population as a function of its timescale relative to
the switching time of the bistable prey population.  

\subsection{Quasi-stationary densities and mean switching times} 

Following the same procedure as for the chemical system, we compute
the stationary density $\rho(x)$ of the prey (in this case it is a
stationary distribution conditioned on the fact that extinction has
not occurred, the so-called quasi-stationary density
\cite{Nasell:2001uc}).\footnote{This is because if we waited long
  enough in the stochastic model, extinction would eventually occur,
  and therefore the stationary density is not defined. In other words,
  there is a ``leak'' at $x=0$.}

\begin{figure*}[htb]
\begin{center}
\includegraphics[height=.3\textwidth]{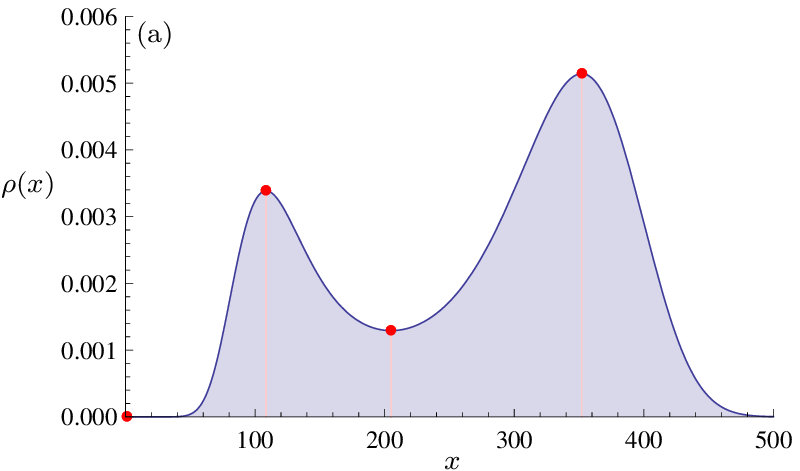} 
\includegraphics[width=.47\textwidth]{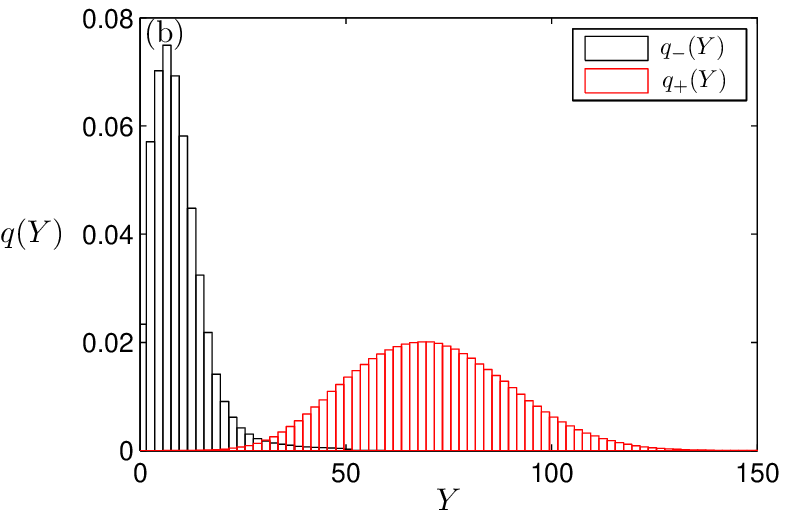} 
\caption{(a) Quasi-stationary marginal density $\rho (x)$ computed from \eqref{stat_x} using the drift and diffusion coefficients in \eqref{driftdiff_prey}. (b) Quasi-stationary marginal densities $q_\pm(Y)$ of $Y$ from SSA of \eqref{reactions_pp} conditional on the prey being at low ($-$, black histogram) and high ($+$, red histogram) levels respectively, obtained from $10^7$ steps of the NRM algorithm. We use $\tau_y = 100$ and the parameter values \eqref{parameters_pp}.} 
\label{fig:Fig16}
\end{center}
\end{figure*}

It is well-known that the Chemical Langevin  approximation should not
be used to predict the extinction rate or the quasi-stationary density
near the extinction state, because it fails to correctly describe the
very large fluctuations necessary to reach the absorbing state of zero
particles.\cite{Kessler:2007ex}  However, here we focus on a
parameter regime for which the metastable prey population is large and
the relaxation time to its quasi-steady density $\rho(x)$ is extremely
small compared to its mean time to extinction. We thus find that the
stationary solution of the Fokker--Planck equation with a reflecting
boundary condition at 
$x=0$, equation \eqref{stat_x}, gives an accurate estimate of the prey
quasi-stationary density. Methods to determine this density more
accurately in the  region near extinction are available (see \eg
 Ref.~\onlinecite{Assaf:2010de}).   
The drift and diffusion coefficients are
\begin{equation}
\label{driftdiff_prey}
v(x) = x [k_1(x)-k_2(x)],\quad  d(x) = \frac{1}{2} x [k_1(x)+k_2(x)],
\end{equation}
The resulting stationary density, which we denote again by $\rho (x)$, is shown in \figref{fig:Fig16}(a). We observe that $\rho(x)$ is bimodal, indicating that the prey population is metastable and will switch between its two favourable states $x_- = 108.5$ and $x_+ = 352.1$.

In \figref{fig:Fig16}(b) we show the histograms of the
quasi-stationary predator distributions $q_\pm(Y)$ conditioned on the
event of low and high prey levels respectively ($X$ lower or higher
than $x_* = 205.0$), for the parameter values in \eqref{parameters_pp} and $\tau_y = 100$. We use the capital letter $Y$ to emphasise that
we are not taking the continuum limit  here (assuming $Y$ is large is
not appropriate if we are interested in 
 extinction)

Next, we evaluate the mean-switching times between the low and high
prey-population levels as in section \ref{sec:mst}.
The mean time $X$ spends in each well is estimated from the stationary
density shown in   \figref{fig:Fig16}(a). The evolution of
the mean of $X$ is calculated by averaging over $10^6$
realisations up to  $T_f = 80$. As before, we fit an
exponential decay to the portion of the curve in which
$\overline{x}-\overline{x}_\infty$ lies between 80\% and 40\% of its
initial value.
From this analysis we obtain the following values for $\theta, \psi$ and $T_\pm$:
\begin{equation}
\label{met_pp}
\begin{aligned}
\theta &= 0.3004,  &  \psi &= 76.8366, \\
 T_-  &= 109.8315, & \quad T_+  &= 255.7689.
\end{aligned}
\end{equation}
We note that the mean switching times $T_\pm$ set the size of the  switching timescale $\tau_s$. 
Having obtained the quasi-steady densities $\rho_\pm(x)$ of the fast
metastable species and its mean switching times $T_\pm$, we are ready
to apply the reduced model approximations to simulate the
predator--prey system. Our goal is to estimate the mean time
to extinction (MTE) of the predator population.  

\subsection{Mean times to extinction} \label{sec:met}

Since we are interested in extinction we do not use a continuum
approximation for $Y$ but retain a discrete approximation.
We denote by $\mathcal T_n$ the MTE of $Y$ given that $Y(0)=n$. It can be
determined exactly via the backward master equation.\cite{gardiner2004handbook} We will use this approach on the original
model as well as on the reduced ones, as detailed below. 

\subsubsection{Full system}
We denote by $\mathcal T_n^m$ the two-dimensional MTE of the full system \eqref{reactions_pp} given that $X(0)= m$ and $Y(0) = n$. It can be easily shown [see analogous one-dimensional example in \eqref{MTE1} below] that it  satisfies 
\begin{align}
\label{MTEfull}
\begin{aligned}
-1 &=  \alpha_1(m) \mathcal T_n^{m+1} + \alpha_2(m) \mathcal T_n^{m-1}   \\
& \phantom{=} +  \alpha_3(n) \mathcal T_{n+1}^m + \alpha_4 (m, n) \mathcal T_{n-1}^m \\
& \phantom{=} - [\alpha_1(m) + \alpha_2(n) + \alpha_3(n) + \alpha_4(m,n)] \mathcal T_n^m,
\end{aligned}
\end{align}
with absorbing left-boundary conditions $\mathcal T_0^m = 0$, $\mathcal T_n^0 = t_n$ and a boundary condition at $m = \kappa$, $\alpha_2(n) (\mathcal T_n^{m-1}- \mathcal T_n^{m}) = 1$, using that $\alpha_1(\kappa) = 0$. Here $t_n$ is the mean  time to extinction when there is no
prey, which can be found by solving Eq. \eqref{MTE1} below
replacing $\overline{\alpha}_4(n)$ by $\alpha_4(0, n)$. In principle this defines an infinite set of difference equations for $\mathcal T_n^m$ for $n\ge 1$ and $0 \le m \le \kappa$. However, noting that $\overline \alpha_4 \gg \alpha_3$ for large $n$ (due to the quadratic term in $\overline \alpha_4$, representing the competition), we can introduce an artificial right boundary condition at $n = N $ for large $N$ and use a similar argument to that of $m=\kappa$ to give $\alpha_4(m, N) ( \mathcal T_{N-1}^m - \mathcal T_N^m)  = -1$. This boundary condition can be imposed by adopting the convention that $\alpha_3(N) = 0$. Finally, to obtain $\mathcal T_n$ we integrate $\mathcal T_n^m$ against the quasi-stationary density $\rho(x)$ in \figref{fig:Fig16}(a).

\subsubsection{Reduced model 1 ($\tau_s \ll \tau_y$)}  
The appropriate reduced model for the system in Regime 1, analogous to RM1 in \eqref{reactiony_reg1}, is
\begin{equation}
\label{predator_reg1}
\begin{aligned}
Y &\to Y+1: \ \alpha_3(Y) = \tilde \lambda Y, \\
Y &\to Y-1: \ \overline \alpha_4(Y)  = \tilde \mu Y +  \frac{\tilde \beta Y^2}{\overline K},  \ \
\frac{1}{\overline K} = \! \int_0^\infty \! \frac{\rho(x)}{K(x)} \, \ud x.
\end{aligned}
\end{equation}
To obtain the MTE from \eqref{predator_reg1}, we consider what can the predator population do in the first short time interval $\delta
t$:
\begin{equation}
\begin{aligned}
\mathcal T_n - \delta t = & \,  \alpha_3(n) \delta t \mathcal T_{n+1} + \overline \alpha_4(n) \delta t \mathcal T_{n-1} \\
& + \left[ 1-  \alpha_3(n) \delta t  - \overline \alpha_4(n) \delta t  \right] \mathcal T_{n}.
\end{aligned}
\end{equation}
Hence we obtain
\begin{equation}
\label{MTE1}
\alpha_3(n) \mathcal T_{n+1} - [\alpha_3(n) + \overline \alpha_4(n)] \mathcal T_n + \overline \alpha_4 (n) \mathcal T_{n-1} = -1.
\end{equation}
Similarly as before, we solve \eqref{MTE1} together with $\mathcal T_0 = 0$ and $\alpha_3(N) = 0$ for $N$ large.  

\subsubsection{Reduced model 2 ($\tau_s \sim \tau_y$)}
Analogously to RM2 defined in \eqref{reactiony_reg2}, the reduced model in Regime 2 has a death rate dependent on whether the prey is at a low or high population level:
\begin{subequations}
\label{predator_reg2}
\begin{equation}
\begin{aligned}
Y \! &\to \! Y+1\!: \, \alpha_3(Y) = \tilde \lambda Y, \\
Y \! &\to \! Y-1\!: \, \alpha_4^\pm (Y)  = \tilde \mu Y +  \frac{\tilde \beta Y^2}{\overline K_\pm},  \ \
\frac{1}{\overline K_\pm} = \! \! \int_{\Omega_\pm} \! \frac{\rho_\pm(x)}{K(x)} \, \ud x,
\end{aligned}
\end{equation}
where $\rho_\pm(x)$ are the quasi-stationary prey densities taken from $\rho(x)$ conditioned on $X\in \Omega_\pm$ (left or right wells). The switches between low- and high-level prey population obey the reactions
\begin{equation}
\label{prey_reg2}
\ce{$S_-$  <=>[k_-][k_+] $S_+$}, \qquad k_\pm = 1/T_\pm 
\end{equation}
\end{subequations}
with $T_\pm$ given in \eqref{met_pp}. This time the MTE must take
into account the initial state of the boolean variable $S(t)$. Denote
$\mathcal T_n^\pm$ the MTE given that $S(0) = S_\pm$ and $Y(0) =n$,
respectively. Then, following a similar argument as before, $\mathcal
T_n^\pm$ obey 
\begin{align}
\label{MTE2}
\begin{aligned}
&\alpha_3(n) \mathcal T^\pm_{n+1} + \alpha_4^\pm (n) \mathcal T^\pm_{n-1} + k_\pm \mathcal T^\mp_n \\
& \quad - [\alpha_3(n) + \alpha_4^\pm(n) + k_\pm] \mathcal T^\pm_n = -1, 
\end{aligned}
\end{align}
with $\mathcal T_0^\pm = 0$ and by convention $\alpha_3(N) = 0$. Again
this results 
in a closed set of equations, now with $2N$ equations and unknowns. 

\subsubsection{Reduced model 3 ($\tau_y \ll \tau_s$)}
Recall that the reduced model 3 does not keep track of explicit $Y$ dynamics but instead only those of the switching variable $S$. In order to extract a MTE in this reduced model, we must introduce a new pair of reactions to represent $Y$-extinction from each of the wells:
\begin{equation}
\label{prey_reg3}
\ce{$S_-$  <=>[k_-][k_+] $S_+$}, \qquad  S_- \stackrel[]{r_-}{\longrightarrow} \emptyset, \qquad S_+ \stackrel[]{r_+}{\longrightarrow} \emptyset,
\end{equation}
where $k_\pm = 1/T_\pm$ as before and
$\emptyset$ means ``$Y$ is extinct''. Here the rates of predator
extinction from the right and left wells are $r_\pm = 1/ \Pi_\pm$,
where $\Pi_\pm$ are the MTE of $Y$ starting from its quasi-stationary
distribution $q_\pm(Y)$ and conditional of $X$ staying in a given well
for all times. To evaluate $\Pi_\pm$, we solve the equation for the MTE in the full system \eqref{MTEfull} but with a reflective boundary condition at $X = x_*$, and integrate the resulting matrix $\mathcal T_n^m$ against the quasi-stationary densities $Q_\pm(m,n)$ (computed similarly as for the chemical system). For any $\tau_y$ one  finds that $\Pi_- \ll \Pi_+$ since it is much more likely for the predator to become extinct when the prey population is in the left well than when it is in the right well [as can be seen in \figref{fig:Fig16}(b)].
From \eqref{prey_reg3} it is easy to show that the MTE $\mathcal T^\pm$ of the reduced model in Regime 3 given that $S(0) = S_\pm $ is 
\begin{equation}
\label{MTE3}
\mathcal T^\pm = \frac{k_-  + k_+ + r_\mp}{ k_+ r_- + k_- r_+ + r_- r_+}.
\end{equation}
The approximation given by RM3 of the MTE $\mathcal T_n$ if we don't know where $X(t)$ started [$X(0) \sim \rho(x)$] is given by $\mathcal T = \theta \mathcal T^- + (1-\theta) \mathcal T^+$. Note that the result in RM3 does not depend on the initial value of $Y$. 

\subsubsection{Results}

\begin{figure}[htb!]
\begin{center}
\includegraphics[width=\columnwidth]{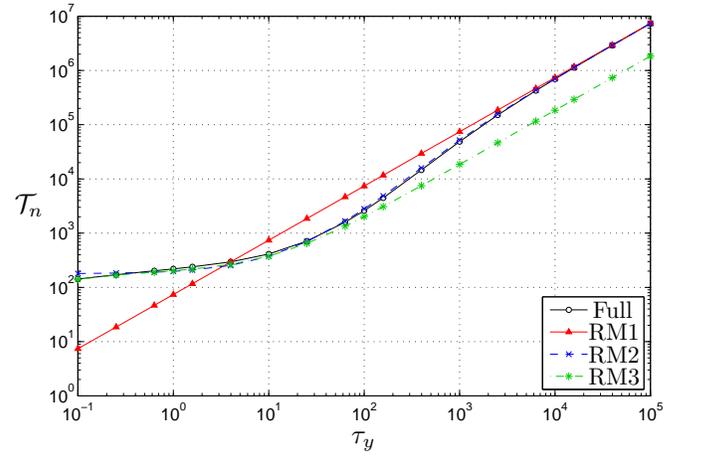} 
\caption{Theoretical mean time to extinction $\mathcal T_n$ of $Y$ as a function of $\tau_y$. Curves obtained from the reduced models RM1 \eqref{MTE1}, RM2 \eqref{MTE2} and RM3 \eqref{MTE3}, and the exact full system MTE \eqref{MTEfull}. We use $n=30$, $N = 150$ and the parameter values \eqref{parameters_pp} and \eqref{met_pp}. The system \eqref{reactions_pp} is in Regime 2 for $\tau_y$ around 100, and moves towards Regime 1 (3) for larger (smaller) values of $\tau_y$.}
\label{fig:Fig17}
\end{center}
\end{figure}

Using the exact expressions defined above for the MTE of $Y(t)$ in the original system as well as in each reduced model, we can look at the accuracy of each reduced model relative to the exact full model as a function of $\tau_y$. We show the results in \figref{fig:Fig17} for $Y(0)=30$ and the parameter values \eqref{parameters_pp} and \eqref{met_pp}. We choose the right boundary at $n=N$ such that the results are insensitive to $N$;  we find that $N = 150$ is a good choice.  The region of validity of each reduced model is set by the switching scale $\tau_s \sim 10^2$ ({\em e.g.}, when $\tau_y \sim 10^2$ the system is in Regime 2). 

\begin{table*}[htdp]
\caption{\label{table3} Mean time to extinction $\mathcal T_{n}$ of $Y$ with $Y(0) = n = 30$ obtained from theory and simulations of the full system and the reduced models in three parameter $\tau_y$ regimes. We use the same parameters and equations as in \figref{fig:Fig17}. For the simulated MTE, we run iterations of the SSA until the standard error in the estimate of $\mathcal T_{n}$ is below $1\%$ and indicate the execution times in parenthesis.} 
\begin{ruledtabular}
\begin{tabular}{ccccccc}
\multirow{2}{*}{Model} & \multicolumn{2}{c }{Regime 1 ($\tau_y = 10^4$)} & \multicolumn{2}{c}{Regime 2 ($\tau_y = 100$)} &  \multicolumn{2}{c}{Regime 3 ($\tau_y = 1$)}   \\
\cline{2-7}
 & Th. & SSA & Th. & SSA  & Th. & SSA \\ \hline  
Full & $6.968\!\times\!10^5$ &
$6.545\!\times\!10^5$ ($\approx$97h)\footnote{Value of $\mathcal T_{30}$ for $\tau_y = 10^4$ computed from 400 runs of \eqref{reactions_pp} (relative error of  6\%). CPU time estimated from the execution time of 400 runs and the $10^4$ runs required to obtain values with $1\%$ relative error (estimated from the number of rounds required for $\tau_y = 10^4$ in the reduced models).} & 2600.5 & 2566.7  (1097s) & 219.24 & 219.58  (464s) \\ 
RM1 & $7.394\!\times\!10^5$ &
$7.313\!\times\!10^5$ (11s) &7394.5 &  7333.2  (11s) & 73.945 & 73.975  (10s)  \\ 
RM2 & $7.043\!\times\!10^5$ & $7.058\!\times\!10^5$
  (13s)& 2790.8 &  2796.9  (8s) & 197.64 & 199.98  (162s)  \\ 
RM3 &$1.840\!\times\!10^5$&   $1.825\!\times\!10^5$  (0.9s) & 2015.3 &  2018.1 (0.01s) & 202.25 &  203.709 (0.005s)   \\
\end{tabular}
\end{ruledtabular}
\end{table*}%
\begin{figure*}[htp]
\begin{center}
\includegraphics[height=.24\textwidth]{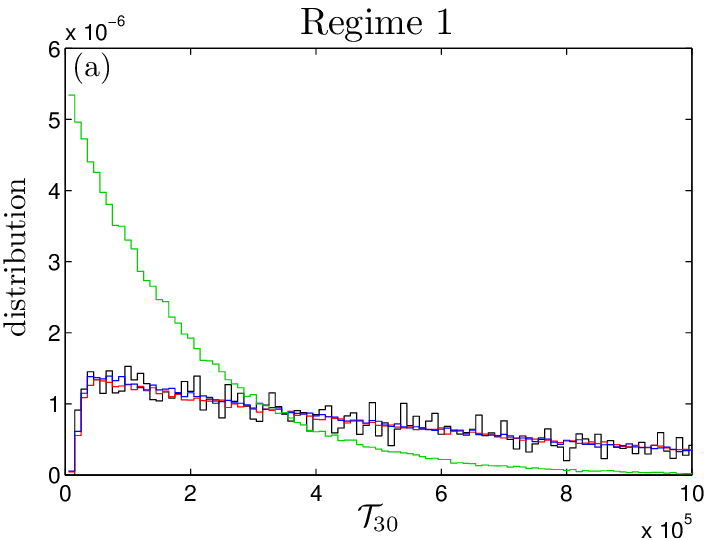} 
\includegraphics[height=.24\textwidth]{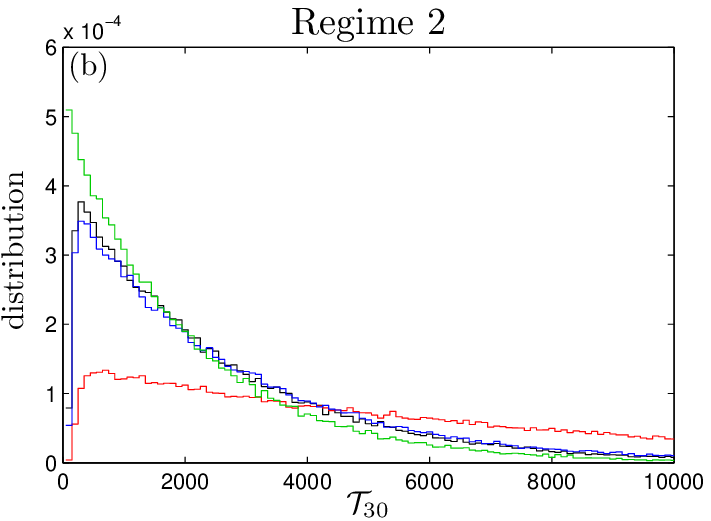} 
\includegraphics[height=.24\textwidth]{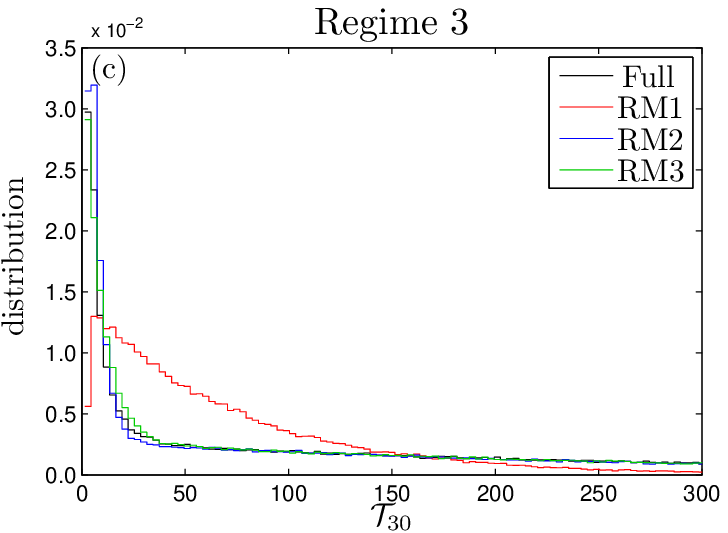} 
\caption{Distribution of the time to extinction $\mathcal T_{n}$ with $Y(0) = n = 30$ for the parameters in \eqref{parameters_pp} and $\tau_y = 10^4, 100$ and 1 (corresponding roughly to a system in Regime 1, 2 and 3 respectively). Histograms computed by $10^5$ rounds of the SSA of the full system \eqref{reactions_pp}, and reduced models 1 \eqref{predator_reg1},  2 \eqref{predator_reg2} and 3 \eqref{prey_reg3}, respectively [except for the full system in (a), in which only 3300 rounds are used].}
\label{fig:Fig18}
\end{center}
\end{figure*}

We find that the MTE computed from the full model increases with $\tau_y$ roughly linearly (as expected) from $\tau_y \sim O(10^2)$, but that for faster timescales it becomes less sensitive to $\tau_y$ (see black line with circles in \figref{fig:Fig17}). This is because, when the system is in Regime 3 and $Y(t)$ is fast, the main contribution to the $\mathcal T$ is the time it takes for a switch in $X$ from the right to the left well to occur ({\em i.e.} $T_+ = 1/ k_+$ on average) multiplied by the probability that $X$ started in the right well, which is $1-\theta$. Once the switch to the low-level prey has occurred, the extinction of $Y$ is almost instantaneous for $\tau_y\to 0$. This is because this extinction occurs at a rate  $r_-$ [see \eqref{prey_reg3}] which scales like $1/\tau_y$. If we do this simple calculation, we find that $\mathcal T_n$ should tend to $T_+ (1-\theta) = 178.94$ as $\tau_y \to 0$, using \eqref{met_pp}. This value is consistent with the results plotted in \figref{fig:Fig17}.
We see that the approximation of the MTE given by RM2 agrees very well with the exact MTE from the full system throughout the whole range of timescales $\tau_y$ except for very small scales where a small error can be perceived. As expected, the approximation to the MTE of RM1 is good for slow timescales $\tau_y$ such that the system is in Regime 1 ($\tau_y \to \infty$), but very poor for fast timescales in Regime 3 when the MTE is highly dependent on the bistable prey behaviour. Conversely, the approximation of RM3 is good in Regime 3 ($\tau_y$ small), but underestimates the MTE of $Y$ when the predator is very slow (and is thus not as sensible to switches to the low-level in the prey population).

In Table \ref{table3} we compare the theoretical results of \figref{fig:Fig17} with those obtained from multiple repetitions of the SSA. We do this to have an idea of the computational time reduction that each reduced model gives relative to the original system. We choose three values of the timescale $\tau_y$ such that they are representative of each regime: $\tau_y = 10^4$ for Regime 1, $\tau_y = 100$ for Regime 2, and $\tau_y =1$ for Regime 3. 
To compute $\mathcal T_n$ numerically, we run the SSA of the full system \eqref{reactions_pp} and the three reduced models RM1 \eqref{predator_reg1}, RM2 \eqref{predator_reg2}, and RM3 \eqref{prey_reg3}. 
 We initialise the system with $Y(0) = 30$ and $X(0) \sim \rho(x)$ [equivalently, for RM2 and RM3,  $S(0)=S_-$ with probability $\theta$] and
 run the simulation until $Y(t)=0$. We record the final time and repeat the process until the standard error in $\mathcal T_n$ is below 1\%. The twelve simulation values of $\mathcal T_{30}$ in  Table \ref{table3} required $O(10^4)$ runs to achieve such relative error. 

Finally, in \figref{fig:Fig18} we show the histograms of the time to extinction $\mathcal T_{30}$ obtained from the simulations. We compare the distribution of $\mathcal T_{30}$ of the full system with those from the reduced models. We see that the Regime 2 model does a good
job of capturing the distribution for the whole range of values of
$\tau_y$. In contrast the reduced models 1 and 3 fail to
capture the distribution of $\mathcal T_{30}$ near the origin
when used outside of their regime of validity. This would be
important if, for instance, we were interested in the probability that
the predator population became extinct within one year.

\section{Summary and discussion} \label{sec:discussion}

In this paper we developed a model reduction technique for slow--fast stochastic systems with metastability. This is a generalisation of previous approaches based on stochastic averaging principles,\cite{Rao:2003gu} when the average of the fast process switches between different quasi-stationary densities instead of relaxing to a unique stationary measure. Interestingly, we found that we can still accurately describe the behaviour of metastable slow--fast systems while improving numerical efficiency substantially by introducing a switch variable to describe the metastable process and using stochastic averaging separately in each of the basins of attraction. 

We focused on a simple class of metastable slow--fast stochastic systems consisting of two species, one of which ($X$) is fast with exponentially long bistable switches. The second species ($Y$), which we assume is the species of interest, has dynamics that are coupled to the bistable species.
We identified three dynamical regimes that led to different model reductions in both of our two-species metastable systems. When the dynamics of $Y$ are much slower than the switches in $X$ (and thus not affected by the bistable behaviour) then the reduced model RM1 is equivalent to the standard fast-variables reduction used in monostable systems.\cite{Rao:2003gu} However, in the other two regimes, when the evolution of $Y$ is of the same order or faster than the switches in $X$, then the switches in $X$ must be taken into account. Our key insight is that two ingredients are required to build a reduced model for slow--fast system with metastability in general. The first ingredient is knowledge of the transition rates between the different basins of attraction for the fast variables (which could of course be more than two in other situations). The second ingredient is the need for estimates of the quasi-stationary densities of the fast variables, conditioned on being in each basin of attraction. This is the procedure  used by Markov state models for the modelling of molecular kinetics, with powerful yet computationally expensive tools available to do this in high-dimensional systems, more suited for parallel computation.\cite{Prinz:2011id}

In the chemical example, we showed how asymptotic approximations of the transition rates can be obtained from the Fokker--Planck equation, as well as from a short SSA run of the full system. In general applications the latter approach is likely to be more feasible because the Fokker--Planck equations can only be used in a limited set of analytically tractable models. However, as seen in section \ref{sec:mst}, care must be taken in defining what it means for the system to have switched to another favourable state. Switching rates have been inferred from other model parameters in a related way for ion channels.\cite{Abad:2009fd,Chen:2014wh} For both the chemical and the ecological examples, we obtained the quasi-stationary densities of the fast variables from both the Fokker--Planck equation equation and the SSA results: again, the limitations of using the Fokker--Planck equation for more general purposes apply. However, once the conditional quasi-steady densities and the switching rates have been obtained, then we can compute the effective rates for the slow variables conditioned on each basin and the transition fluxes between sub-models. 

In this work we assumed a known slow--fast model with fixed parameters. One natural extension to our study is to systems in which the  parameters affecting the regime may change through time. For example, if $\tau_y$ changed through time, then one may be inclined to dynamically change the reduced model being employed. This would require building a set of rules or boundaries to discern between regimes (precomputed), and checking during a simulation whether any of these have been crossed. However, the extra computational effort involved in such a process is likely to be more costly than simply using the reduced model 2 (RM2) throughout the simulation. This model interpolates between Regimes 1 and 3 and is valid in the whole parameter regime for $\tau_y$ (\eg see  \figref{fig:Fig17}). 
To our knowledge this is the first method that can be used in the whole spectrum of $\tau_y$ in the class of systems considered here, thus extending the method in Ref.~\onlinecite{Cao:2006hq} to cases with more than one stable state, while still keeping it simple (a one-dimensional model plus the switch variable), in contrast to more complicated fully two-dimensional models appropriate in Regime 3.\cite{Bressloff:2013ve} A potentially more interesting situation arises if the (quasi-)stationary density of the fast variable $\rho(x)$ changes dynamically; maybe even through a two-way coupling with $y$. In this case, the RM2 as presented would need to be modified to since one needs to update dynamically the mean switching times $T_\pm$ and the conditional densities $\rho_\pm(x)$. Moreover, it could happen that $\rho(x)$ evolved from a  bimodal to unimodal  shape, in which case the system would move into into the monostable regime. A possible strategy would be to establish a timestep $\Delta_y$ such that $\epsilon \ll  \Delta_y \ll \tau_y$, and every $\Delta_y$ stop the simulation, do a short run of the full dynamics to re-evaluate $T_\pm^{\hat y}$ and $\rho^{\hat y}_\pm(x)$, parameterised by current value of $Y= \hat y$. 

We applied our techniques to a metastable system with an absorbing state; the predator--prey stochastic model with bistable prey ($X$), and showed that the reduced models can be used to predict the mean times to extinction (MTE) of the predator population $Y$ in an accurate and efficient way. Fundamentally, this showed that the method can be applied more generally to two-dimensional systems, even with absorbing states. This implies that the specific forms of the rates and the coupling between $X$ and $Y$ do not matter for the application of our model reduction techniques. Moreover, the model reduction has provided some new insights into how the routes to extinction of the predator change as we varied its relative timescale to the prey switching and its quasi-stationary density changed from unimodal to bimodal. In particular, the reduced model RM2  can identify the huge difference between the MTE of $Y$ when the prey population $X$ is at its low or high level, and, most importantly, how this fact together with the bistable process translates into the overall MTE. 
The specific application of our approximations to enable the efficient calculation of the mean time to extinction in Regime 2 is, to our knowledge, new to the literature and clearly illustrates the value of the approximations of reduced model 2 (explicit $Y$ with a switch variable for $X$) over the original system to estimate the mean time to extinction.
 It was fortunate for these equations that we could assess our estimates with the backward master equation; however, this will rarely be possible for higher-dimensional ecological models. It is in such situations that having insight into the appropriate reduction method is most valuable, giving one the ability to select the appropriate stochastic reduced model to obtain accurate approximations efficiently. In high dimensional systems with several variables displaying intrinsic metastable behaviour, one could use a combination of our method presented here for the MTE and the computational approaches discussed in Ref.~\onlinecite{Prinz:2011id} to split the space into basins of attraction and approximate the transitions times.

Finally, this work could be extended to enable long-term predictions of complex real bistable systems. This could, for example, enable new insights into our understanding of critical transitions and our ability to predict them.\cite{Scheffer:2012ct} In recent years there has been a lot of effort in investigating time-series of systems prone to critical transitions in an attempt to enable predictions of transitions.\cite{Dakos:2012ft} 
Could we detect the dynamical regime of a bistable system from its dynamics without knowing the parameters or model? If we could then it is possible to imagine using similar ingredients to those used here to infer the mean switching times characterising the metastable processes and the reaction time (after a switch) for the variables of interest. 

\begin{acknowledgments}
We thank J.~M. Newby, G.~M. Palamara, and Y.~G. Kevrekidis for useful discussions on aspects of this article. We also thank M. Geissbuehler and T. Lasser for their \emph{Morgenstemning} colormap.\cite{Geissbuehler:2013er} M.B. is partially funded by the EPSRC (EP/I017909/1) and Microsoft Research, Cambridge and by St John's College, Oxford,  in the form of a Junior Research Fellowship.
\end{acknowledgments}

\appendix

\section{Perturbation analysis of the three bistable regimes} \label{sec:perturbation}

In this appendix we provide formal derivations of the three reduced models in the main text using a perturbation analysis at the level of the Fokker--Planck equation for the joint probability density $P(x, y, t)$. 
In \ref{sec:regime1} we consider the asymptotic regime $\tau_s \ll \tau_y$ (Regime 1), resulting in the reduced model 1. The more interesting reduced model 2, appropriate when $\tau_s \sim \tau_y$, is derived in \ref{sec:regime2} using a WKB perturbation method. Finally, in \ref{sec:regime3} we discuss the regime when $\tau_y \ll \tau_s$ and the reduced model 3 is appropriate. 

We consider the Fokker--Planck equation \eqref{fokker-planck} for the joint
probability density $P(x,y,t)$.
The main timescale of interest is that of the evolution of $Y$, so let us
rescale time with $\tau_y$ to give
\begin{equation}
\label{fokker-planck_rescaled}
\begin{aligned}
\frac{\partial P}{\partial t} (x,y,t) = \, & \frac{\tau_y}{\epsilon}
\frac{\partial}{\partial x} \left\{  \frac{\partial}{\partial x} [
  d(x) P] - v(x) P \right\} \\
  &+  \frac{\partial}{\partial y} \left\{
  \frac{\partial}{\partial y} [ D(x,y) P] - V(x,y) P \right\},
  \end{aligned}
\end{equation}
where $\tau_y/\epsilon\gg 1$.

\subsection{Regime 1: $\epsilon \ll \tau_s \ll \tau_y$}
\label{sec:regime1}

First we consider equation \eqref{fokker-planck_rescaled} for short
times such that $t = O(\epsilon)$. We define the fast time
$\tilde t$ as $t = (\epsilon/\tau_y) \tilde t$ and write $\tilde P(x, y, \tilde
t) = P(x, y, t)$, to give, at leading order,
\begin{equation}
\label{fp_reg10}
\frac{\partial \tilde P^{(0)}}{\partial \tilde t}  = \frac{\partial}{\partial x} \left\{  \frac{\partial}{\partial x} [ d(x) \tilde P^{(0)}] - v(x) \tilde P^{(0)} \right\}.
\end{equation}
As  $\tilde t \to \infty$, the solution of \eqref{fp_reg10} converges
to 
\begin{equation}
\label{reg1_ss}
\tilde P^{(0)} (x, y) = \rho (x) q(y) , 
\end{equation}
where $\rho(x)$ is the normalised steady solution of
\eqref{fp_reg10} [given by \eqref{stat_x}], and 
\[ q(y) = \int_0^\infty P(x,y,0)\, {\rm d} x.\]
 Now we move
back to $O(1)$ times and consider equation
\eqref{fokker-planck_rescaled}.
Expanding $P \sim P^{(0)} + \epsilon/\tau_y \, P^{(1)} + \cdots$,
gives, at leading-order,
\begin{equation}
\label{fp_reg11}
\frac{\partial}{\partial x} \left\{  \frac{\partial}{\partial x} [
  d(x)  P^{(0)}] - v(x)  P^{(0)} \right\} = 0.
\end{equation}
Thus
\begin{equation}
\label{reg1_lo}
P^{(0)}(x, y, t) = C(y, t) \rho(x), 
\end{equation}
where $C(y, t)$ is arbitrary at this stage. Matching this solution for
long times with the short times solution \eqref{reg1_ss} gives  $C(y, 0)= q(y)$. At the next order equation 
\eqref{fokker-planck_rescaled} gives 
\begin{equation}
\label{fp_reg11_o1}
\begin{aligned}
\frac{\partial P^{(0)}}{\partial t} =\, &\frac{\partial}{\partial x} \! \left\{  \frac{\partial}{\partial x} [ d(x) P^{(1)}] - v(x) P^{(1)} \right\} \\
&+  \frac{\partial}{\partial y}\! \left\{  \frac{\partial}{\partial y} [ D(x,y) P^{(0)}] - V(x,y) P^{(0)} \right\}.
\end{aligned}
\end{equation}
Integrating this equation with respect to $x$ (using no-flux boundary
conditions at $x = 0$, $\infty$) gives the following solvability
condition for $C$: 
\begin{align}
\label{solva0}
\frac{\partial C}{\partial t} (y,t)  = \frac{\partial}{\partial y} \! \left\{  \frac{\partial}{\partial y} \left[  \overline D(y) C(y, t)\right] -  \overline V(y) C(y, t) \right\}
\end{align}
where
\[
\overline F(y) = \int F(x, y) \rho(x) \, \ud x, \qquad \mbox{ for }F = D, \ V.
\]
Finally, since
\[ C(y,t) = \int_0^\infty P(x,y,t)\, {\rm d}x\]
we see that $C(y,t)$ is the marginal density for $Y(t)$, so that
\eqref{solva0} gives the evolution of the probability density function
of a reduced process for $Y(t)$ where the fast variable 
$X(t)$ has been averaged out. Going back to the original time variable we can then write
\begin{align}
\label{solva}
\frac{\partial p}{\partial t} (y,t)  = \frac{1}{\tau_y} \frac{\partial}{\partial y} \left\{  \frac{\partial}{\partial y} \left[  \overline D(y) p \right] -  \overline V(y) \right\}
\end{align}
Equation \eqref{solva} is the Fokker--Planck equation associated with the reduced stochastic model  \eqref{reactiony_reg1} appropriate for Regime 1 when $Y$ is approximated by a continuous random variable.   
This asymptotic reduction relies on the fact that the stochastic
process for $X$ reaches steady state on a timescale which is faster
than the timescale for the evolution of $Y$.

\subsection{Regime 2: $\epsilon \ll \tau_s \sim \tau_y$}
\label{sec:regime2}
  To analyse Regime 2 asymptotically we
need to ensure that $X$ is metastable, and quantify the switching
time. We suppose then that $1/\delta$, the typical equilibrium value of
$X$, is large, and scale the rate constants as in \eqref{values}. We
set $\hat{x} = \delta x$, where $\hat{x}$
is $O(1)$ as $\delta \to 0$. We will see that 
$\tau_s \sim e^{r/\delta}$ for some constant $r>0$.  In terms of the
new scaled variables equation
\eqref{fokker-planck_rescaled} becomes
\begin{equation}
\label{fokker-planck_reg2}
\begin{aligned}
\frac{\partial P}{\partial t} = \, &\frac{ \tau_y}{\epsilon}
\frac{\partial}{\partial \hat{x}} \left\{  \frac{\partial}{\partial \hat{x}} [
  \delta \hat{d}( \hat{x}) P] - \hat{v}( \hat{x}) P \right\} \\
  & +
\frac{\partial}{\partial y} 
\left\{  \frac{\partial}{\partial y} [ D(\hat{x},y) P] - V(\hat{x},y) P \right\},
\end{aligned}
\end{equation}
where $\hat{d}(\hat{x}) = \delta d(x)$, etc.
With this scaling we see that the diffusion $\delta \hat{d}(\hat{x})$
of $X$ is weaker than the drift $\hat{v}(\hat{x})$ when $\delta \ll 1$, 
which is the reason
 that switches do not occur frequently  ($\tau_s \gg \epsilon$); the
 parameter $\delta$ makes explicit this separation of timescales
 between drift of $X$ and switches in $X$ (this is the weak noise limit\cite{bressloff:2013wn}).

It is convenient to write \eqref{fokker-planck_reg2} in the form
\begin{equation}
\label{fp_op}
\hat{\epsilon} \frac{\partial P}{\partial t} + \hat{\epsilon}
\frac{\partial J_y}{\partial y} = \mathcal L_\delta, 
\end{equation}
where $\hat{\epsilon} = \epsilon/\tau_y$, $J_y =
-\frac{\partial}{\partial y} [ D(\hat{x},y) P] + V(\hat{x},y) P$ and $\mathcal
L_\delta$ is a linear operator acting on the variable $\hat{x}$ only:   
\begin{equation}
\label{fp_reg2_0}
\mathcal L_\delta P \equiv \frac{\partial}{\partial \hat{x}} \left (  \frac{\partial}{\partial \hat{x}} [ \delta \hat{d}(\hat{x}) P] - \hat{v}(\hat{x}) P \right ).
\end{equation}
The marginal stationary density in $\hat{x}$ [$\rho(x)$ in
\eqref{stat_x}] corresponds to the zero-eigenvalue 
eigenfunction $\mathcal L_\delta$, that is
$\mathcal L_\delta \rho(\hat{x}) = 0$,  
giving, in the scaled variables,
\begin{equation}
\label{zero_eigenfunction}
\rho (\hat{x}) = \frac{A}{\hat d(\hat{x})} \exp \left [ \frac{1}{\delta}
  \int_0^{\hat{x}} \frac{\hat{v}(s)}{\hat{d}(s)}\, \ud s \right], 
\end{equation}
where $A$ is the normalisation constant.

The novelty in this bistable problem is the coupling of $X$ with $Y$, that
is, in the term $J_y$ in \eqref{fp_op}.
We sketch the following calculation since it follows closely that of Refs.~\onlinecite{Ward:1998ve} and \onlinecite{Hinch:2005eg}.
By using a WKB approximation for small $\delta$ (with $\hat{\epsilon} \ll
\delta$) we find  solutions to \eqref{fp_op} of the form 
\begin{widetext}
\begin{align}
\label{WKBsolution} 
P  \sim  \begin{cases} A_- \phi(\hat x) & \hat x< \hat x_*^v,\\
\frac{1}{2 }  \phi(\hat x_*^v) e^{-\gamma_* (\hat x-\hat x_*^v)^2 / 2\delta} \left [ A_+ + A_- + (A_+ - A_-) \text{erf} \left( \sqrt{ \frac{|\gamma_*|}{2\delta}} (\hat x-\hat x_*^v) \right)  \right]
& \hat x \approx \hat x_*^v,\\ 
A_+ \phi(\hat x) & \hat x>\hat x_*^v. 
\end{cases}
\end{align}
\end{widetext}
where 
\begin{equation}
\label{def_phi}
\phi(\hat x) = \frac{e^{u(\hat x)/\delta}}{\hat d(\hat x)}, \quad
u(\hat x) = \int_0^{\hat x} \frac{\hat v(s)}{ \hat d(s)} \ud s,
\quad \gamma_* = -\frac{\hat v'(\hat x_*^v)}{ \hat d(\hat x_*^v)},
\end{equation}
and $\hat x_*^v$ is the turning point at which $u'(\hat x_*^v)=0$.
Here $A_-$ and $A_+$ are independent of $\hat x$ but are undetermined, and
may depend on both $y$ and 
$t$. Define  
\begin{equation}
\label{Tpm_def}
\Phi_\pm = \frac{1}{A_\pm} \int_{\Omega_\pm} P_\pm \,\ud \hat x \equiv \int_{\Omega_\pm} \phi \,\ud \hat x.
\end{equation}
We note that $A_+ \Phi_+ = 1- \theta$ introduced in section \ref{sec:mst}. These integrals can be evaluated using Laplace's method, giving 
\begin{equation}
\label{Tpm_result}
\Phi_\pm \sim \sqrt{\frac{2\pi \delta}{\gamma_\pm}}   \phi(\hat x_\pm^v), 
\end{equation}
where $\hat x_{\pm}^v$ are the maxima of $u$ (or the zeros of $\hat v$) in $\Omega_{\pm}$ and
\[\gamma_\pm = -\frac{\hat v'(\hat x_\pm^v)}{\hat d(\hat x_\pm^v)}.\]
In the one-dimensional case we could now use the normalisation
condition on $P$ to give a relationship between $A_+$ and $A_-$,
namely  $1 = \Phi_- A_- + \Phi_+ A_+$. But in the two-dimensional case we only have
that 
\begin{align}
\label{def_A}
\begin{aligned}
p(y, t) &= \! \int_0^\infty\! \!P(\hat x,y, t) \, \ud \hat x = A_- (y, t) \Phi_- + A_+(y, t) \Phi_+.
\end{aligned}
\end{align}
 Define
\begin{equation}
\label{marginaly_decoupled}
p_\pm (y, t) = \Phi_\pm A_\pm (y,t),
\end{equation}
which we can interpret as the marginal density for $Y$ given $X$ is in
the left/right well, multiplied by the probability that $X$ is in that
well. To calculate the exponentially slow transition rates we need to
calculate the first eigenvalue/eigenfunction of $\mathcal L_\delta$,
given by
\begin{equation}
\label{first_eig}
\mathcal L_\delta \rho_1 = \lambda_1 \rho_1,
\end{equation}
say.
Following Ward,\cite{Ward:1998ve} a good approximation of the first
eigenfunction $\rho_1$ is the derivative of \eqref{WKBsolution} with
respect to the unknown constant. Using $\rho_1(\hat x)  = \partial
P/ \partial A_-$, gives
\begin{widetext}
\begin{align}
\rho_1(\hat x) \sim  \begin{cases} \phi(\hat x) & \hat x< \hat x_*^v,\\
\frac{1}{2 }  \phi(\hat x_*^v) e^{-\gamma_* (\hat x-\hat x_*^v)^2 / 2\delta} \left [ 1  -\frac{\Phi_-}{\Phi_+} - \left(1+ \frac{\Phi_-}{\Phi_+} \right)  \text{erf} \left( \sqrt{ \frac{|\gamma_*|}{2\delta}} (\hat x-\hat x_*^v) \right)  \right]
& \hat x \approx \hat x_*^v,\\ 
-\frac{\Phi_-}{\Phi_+} \phi(\hat x) & \hat x>\hat x_*^v. 
\end{cases}
\end{align}
\end{widetext}

To obtain $\lambda_1$ we use a spectral projection method that makes
use of the adjoint operator $\mathcal L^*_\delta$,
given by
\begin{equation}
\label{adjoint}
\mathcal L_\delta ^* \varphi \equiv  \left (  \delta \hat d(\hat x)  \frac{\partial^2 \varphi}{\partial \hat x^2}  +  \hat v(\hat x)  \frac{\partial \varphi}{\partial \hat x} \right ),
\end{equation}
together with boundary conditions $\varphi'(\hat x) = 0$ on $\hat x = 0, \infty$. 
The eigenfunctions of the adjoint operator satisfy 
\begin{equation}
\label{adjoint_def}
\mathcal L^*_\delta \xi_j = \lambda_j \xi_j,
\end{equation}
with the orthogonality relationship 
$\langle \rho_i, \xi_j \rangle  = \delta_{ij}$ where $\delta_{ij}$
is the Kronecker delta.
The adjoint eigenfunction corresponding to $\lambda_0 = 0$ is simply
$\xi_0 = 1$. The first adjoint eigenfunction $\xi_1(\hat x)$ is
approximately (see
e.g. Ref.~\onlinecite{Hinch:2005eg})
\begin{align}
\label{adjoint_1}
\xi_1(\hat x) \! \sim \! \begin{cases} \frac{\Phi_+}{\Phi_- (\Phi_+ + \Phi_-)}  & \hat x< \hat x_*^v,\\
\frac{\Phi_+}{\Phi_- (\Phi_+ + \Phi_-)} - 
\frac{1}{2 \Phi_- }  \text{erf} \left( \!\! \sqrt{ \frac{|\gamma_*|}{2\delta}}
  (\hat x-\hat x_*^v) \! \right)  \! \!
& \hat x \approx \hat x_*^v, \\ 
 \frac{-1}{(\Phi_+ + \Phi_-)}  & \hat x>\hat x_*^v,
\end{cases}
\end{align}

The first eigenvalue can now be computed by taking the inner product
of (\ref{first_eig}) with a suitable test function $\varphi$:
\begin{equation}
\langle \varphi, \mathcal L_\delta \rho_1 \rangle = \lambda_1 \langle  \varphi, \rho_1 \rangle.
\end{equation}
Here we use $\varphi = \mathds{1}_{\Omega_-}$. The left hand side gives  
\begin{align}
\label{lhs}
\begin{aligned}
\langle \varphi, \mathcal L_\delta \rho_1 \rangle &= \! \int_0^{\hat x_*^v} \! \!\mathcal L_\delta \rho_1 \, \ud \hat x =  \! \left (  \frac{\partial}{\partial \hat x} [ \delta \hat d(\hat x) \rho_1] - \hat v(\hat x) \rho_1\! \right )\!  \bigg | _{\hat x = \hat x_*^v} \\
&= \sqrt \delta \hat d(\hat x_*^v)  \rho_1'(z) |_{z = 0}\\
&= - \sqrt \delta e^{u(\hat x_*^v)/\delta} \left( 1 + \frac{\Phi_-}{\Phi_+} \right) \sqrt{ \frac{|\gamma_*|}{2 \pi}}.
\end{aligned}
\end{align}
The right-hand side is
\begin{equation}
\label{rhs}
\langle \varphi, \rho_1 \rangle = \int_0^{\hat x_*^v}  \rho_1 \, \ud \hat x = \Phi_- \sim \sqrt{\frac{2 \pi \delta}{\gamma_-}} \phi(\hat x_-^v),
\end{equation}
using \eqref{Tpm_result}. Combining \eqref{lhs} and \eqref{rhs} gives the exponentially small eigenvalue
\begin{equation}
\label{lambda1}
\lambda_1 \sim - \frac{\hat d(\hat x_-^v) }{2 \pi} \left( 1 + \frac{\Phi_-}{\Phi_+} \right)   \sqrt{ \frac{|\gamma_*|}{\gamma_-}} \exp \left( \frac{u(\hat x_*^v) - u(\hat x_-^v)}{\delta} \right).
\end{equation}
Note that the argument of the exponential is negative since $u(\hat x_*^v) < u(\hat x_-^v)$.

Finally, we seek the differential equations describing the evolution
of $A_+$ and $A_-$.
First we integrate the equation \eqref{fp_op} with respect to $\hat x$
(equivalent to taking the inner product  with the adjoint
eigenfunction $\xi_0 = 1$): 
\begin{equation}
\label{proj_0}
\hat{\epsilon} \left \langle 1, \frac{\partial P}{\partial t} +
  \frac{\partial J_y}{\partial y}  \right \rangle = \langle \xi_0,
\mathcal L _\delta P \rangle = \langle \mathcal L_\delta^* \xi_0,
P \rangle = 0. 
\end{equation}
 From the left-hand side  we have
\begin{equation}
\label{proj0time}
\left \langle 1, \frac{\partial P}{\partial t}  \right \rangle = \Phi_-
\frac{\partial A_-}{\partial t} + \Phi_+ \frac{\partial A_+}{\partial t}, 
\end{equation}
and
\begin{equation}
\label{proj0y}
\begin{aligned}
& \left \langle 1, \frac{\partial J_y}{\partial y}  \right \rangle\\
& \quad = \int_{\Omega_-} \frac{\partial}{\partial y} \left (    V(\hat x,y) A_- - \frac{\partial}{\partial y} [ D(\hat x,y) A_-] \right ) \phi(\hat x) \, \ud \hat x \\
& \quad \phantom{=} \ + \! \int_{\Omega_+} \frac{\partial}{\partial y} \left (    V(\hat x,y) A_+ - \frac{\partial}{\partial y} [ D(\hat x,y) A_+]\right ) \phi(\hat x) \, \ud \hat x.
\end{aligned}
\end{equation}
Taking the integrals over $\Omega_\pm$ inside the $y$ derivatives and
using that $\rho_\pm(\hat x) = \phi(\hat x)/\Phi_\pm$, we find
\begin{equation}
\label{proj0y2}
\begin{aligned}
\left \langle \! 1, \frac{\partial J_y}{\partial y}  \! \right \rangle
= \ &\Phi_- \frac{\partial}{\partial y} \! \left (   \overline V_-(y) A_- -
  \frac{\partial [ \overline D_-(y) A_-]}{\partial y}  \right )  \\
  & + \Phi_+
\frac{\partial}{\partial y} \! \left (  \overline V_+(y) A_+ -
  \frac{\partial  [ \overline D_+(y) A_+]}{\partial y}\right ), 
  \end{aligned}
\end{equation}
where
\begin{equation}
\label{effective rates}
\overline F_\pm (y) : = \int_{\Omega_\pm} F(\hat x,y)\rho_\pm (\hat x) \, \ud \hat x,
\quad \mbox{for }F = D, \ V.
\end{equation}
Thus
\begin{align}
\label{proj0_res}
\begin{aligned}
 0 = \  &\Phi_- \frac{\partial A_-}{\partial t} + \Phi_+ \frac{\partial
   A_+}{\partial t} \\ 
& +  \Phi_- \frac{\partial}{\partial y} \! \left (  \overline V_-(y) A_- -  \frac{\partial [ \overline D_-(y) A_-]}{\partial y} \right ) \\
& + \Phi_+ \frac{\partial}{\partial y} \! \left (  \overline V_+(y) A_+ - \frac{\partial  [ \overline D_+(y) A_+]}{\partial y} \right ).
\end{aligned}
\end{align}
Next we take the inner product of \eqref{fp_op} with the first adjoint eigenfunction:
\begin{equation}
\hat{\epsilon} \left \langle \xi_1, \frac{\partial P}{\partial t} + \frac{\partial J_y}{\partial y}  \right \rangle = \langle \xi_1, \mathcal L _\delta P \rangle = \langle \mathcal L_\delta^* \xi_1, P \rangle = \lambda_1 \langle \xi_1, P \rangle.
\end{equation}
Using \eqref{WKBsolution} and \eqref{adjoint_1} gives another PDE for
$A_-$ and $A_+$: 
\begin{align}
\label{proj1_res}
\begin{aligned}
\frac{\partial A_-}{\partial t} & -  \frac{\partial A_+}{\partial t}  -  \frac{\partial}{\partial y} \! \left (  \frac{\partial [ \overline D_-(y) A_-]}{\partial y}  - \overline V_-(y) A_- \right ) \\
 +  &\frac{\partial}{\partial y} \! \left (  \frac{\partial  [ \overline D_+(y) A_+]}{\partial y} - \overline V_+(y) A_+ \right )\\
&  = \frac{\lambda_1}{\hat{\epsilon}} (A_- - A_+).  
\end{aligned}
\end{align}
Rearranging  \eqref{proj0_res} and \eqref{proj1_res} we find the
following system for $A_-(y,t)$ and $A_+(y,t)$: 
\begin{subequations}
\label{final_A}
\begin{align}
\begin{aligned}
&\frac{\partial A_-}{\partial t} -  \frac{\partial}{\partial y} \! \left (  \frac{\partial [ \overline D_-(y) A_-]}{\partial y}  - \overline V_-(y) A_- \right ) \\
&\quad  =\frac{\lambda_1}{\hat{\epsilon}} \frac{\Phi_+}{\Phi_- + \Phi_+} (A_- - A_+),
\end{aligned}\\
\begin{aligned}
&\frac{\partial A_+}{\partial t} - \frac{\partial}{\partial y} \! \left (  \frac{\partial  [ \overline D_+(y) A_+]}{\partial y} - \overline V_+(y) A_+ \right )\\
& \quad = \frac{\lambda_1}{\hat{\epsilon}} \frac{\Phi_-}{\Phi_- + \Phi_+} (A_+ - A_-).
\end{aligned}
\end{align}
\end{subequations}
Using \eqref{marginaly_decoupled}, we can write \eqref{final_A}
in terms of probabilities, $p_\pm(y, t)$:
\begin{subequations}
\label{final_p}
\begin{align}
\begin{aligned}
&\frac{\partial p_-}{\partial t} -  \frac{\partial}{\partial y} \! \left (  \frac{\partial [ \overline D_-(y) p_-]}{\partial y}  - \overline V_-(y) p_- \right ) \\
& \quad =\frac{\lambda_1}{\hat \epsilon} \frac{1}{\Phi_- + \Phi_+} (\Phi_+ p_- -\Phi_- p_+),
\end{aligned}
\\
\begin{aligned}
&\frac{\partial p_+}{\partial t} -  \frac{\partial}{\partial y} \! \left (  \frac{\partial  [ \overline D_+(y) p_+]}{\partial y} - \overline V_+(y) p_+ \right )\\
&\quad  = \frac{\lambda_1}{\hat \epsilon} \frac{1}{\Phi_- + \Phi_+} (\Phi_-p_+ - \Phi_+p_-).
\end{aligned}
\end{align}
\end{subequations}
Finally, rescaling time in \eqref{final_p} back to the original time variable [recall that in \eqref{fokker-planck_rescaled} we had scaled time with $\tau_y$], we find that 
\begin{subequations}
\label{final_p2}
\begin{align}
\frac{\partial p_-}{\partial t} -  \frac{1}{\tau_y}\frac{\partial}{\partial y} \! \left \{  \! \frac{\partial [ \overline D_-(y) p_-]}{\partial y}  - \overline V_-(y) p_- \! \right 
\} \! & =k_+ p_+ \! - \! k_- p_- ,\\
\frac{\partial p_+}{\partial t} -  \frac{1}{\tau_y}\frac{\partial}{\partial y} \! \left \{ \!  \frac{\partial  [ \overline D_+(y) p_+]}{\partial y} - \overline V_+(y) p_+ \!\right \}\! & = k_- p_- \! - \! k_+ p_+,
\end{align}
\end{subequations}
where
\begin{equation}
k_- = \frac{\Phi_+}{\Phi_- + \Phi_+} \frac{|\lambda_1| }{\epsilon}, \qquad k_+ = \frac{\Phi_-}{\Phi_- + \Phi_+} \frac{|\lambda_1| }{\epsilon}.
\end{equation}
These rates can be identified as the transition rates from the left to right
well and vice versa (introduced in section \ref{sec:mst}). 

Equations \eqref{final_p2} are the Fokker--Planck equations associated
with the reduced stochastic model \eqref{reactiony_reg2} appropriate
for Regime 2 when $Y$ is approximated by a continuous random variable.

\subsection{Regime 3: $\epsilon \sim \tau_y \ll \tau_s$}
\label{sec:regime3}

In Regime 3 both $X$ and $Y$ switch between localised metastable
states.
A similar analysis to that in \S\ref{sec:regime2} can be used. 
In our simplified example in which the bistable variable $X$ is
independent of $Y$ the switching rate is exactly given by
\S\ref{sec:regime2}; all that remains is to calculate the
quasi-stationary density  for each metastable state.
To approach this analytically requires a two-dimensional WKB (ray
theory) approach, which is considerable more complicated than the
one-dimensional version in \S\ref{sec:regime2}.

If the system was fully coupled, a two-dimensional version of the
eigenvalue calculation of \S\ref{sec:regime2} would be required to
analytically determine the transition rates.\cite{Bressloff:2013ve}
Alternatively we could  numerically obtain the
quasi-stationary densities of each
well and the mean switching times  between attractors using short
bursts of stochastic simulation, as described in Sec. \ref{sec:mst}.
Such a numerical approach could in principle be  extended to an
arbitrary number of attractors and/or higher dimensions as an
automated process. However, as we have seen, care needs to be taken to
define the 
boundaries between attractors and in determining the switching times.\cite{Hartmann:2013wi} 


\begin{thebibliography}{50}%
\makeatletter
\providecommand \@ifxundefined [1]{%
 \@ifx{#1\undefined}
}%
\providecommand \@ifnum [1]{%
 \ifnum #1\expandafter \@firstoftwo
 \else \expandafter \@secondoftwo
 \fi
}%
\providecommand \@ifx [1]{%
 \ifx #1\expandafter \@firstoftwo
 \else \expandafter \@secondoftwo
 \fi
}%
\providecommand \natexlab [1]{#1}%
\providecommand \enquote  [1]{``#1''}%
\providecommand \bibnamefont  [1]{#1}%
\providecommand \bibfnamefont [1]{#1}%
\providecommand \citenamefont [1]{#1}%
\providecommand \href@noop [0]{\@secondoftwo}%
\providecommand \href [0]{\begingroup \@sanitize@url \@href}%
\providecommand \@href[1]{\@@startlink{#1}\@@href}%
\providecommand \@@href[1]{\endgroup#1\@@endlink}%
\providecommand \@sanitize@url [0]{\catcode `\\12\catcode `\$12\catcode
  `\&12\catcode `\#12\catcode `\^12\catcode `\_12\catcode `\%12\relax}%
\providecommand \@@startlink[1]{}%
\providecommand \@@endlink[0]{}%
\providecommand \url  [0]{\begingroup\@sanitize@url \@url }%
\providecommand \@url [1]{\endgroup\@href {#1}{\urlprefix }}%
\providecommand \urlprefix  [0]{URL }%
\providecommand \Eprint [0]{\href }%
\providecommand \doibase [0]{http://dx.doi.org/}%
\providecommand \selectlanguage [0]{\@gobble}%
\providecommand \bibinfo  [0]{\@secondoftwo}%
\providecommand \bibfield  [0]{\@secondoftwo}%
\providecommand \translation [1]{[#1]}%
\providecommand \BibitemOpen [0]{}%
\providecommand \bibitemStop [0]{}%
\providecommand \bibitemNoStop [0]{.\EOS\space}%
\providecommand \EOS [0]{\spacefactor3000\relax}%
\providecommand \BibitemShut  [1]{\csname bibitem#1\endcsname}%
\let\auto@bib@innerbib\@empty
\bibitem [{\citenamefont {Thomas}, \citenamefont {Straube},\ and\ \citenamefont
  {Grima}(2012)}]{Thomas:2012ej}%
  \BibitemOpen
  \bibfield  {author} {\bibinfo {author} {\bibfnamefont {P.}~\bibnamefont
  {Thomas}}, \bibinfo {author} {\bibfnamefont {A.~V.}\ \bibnamefont {Straube}},
  \ and\ \bibinfo {author} {\bibfnamefont {R.}~\bibnamefont {Grima}},\
  }\href@noop {} {\bibfield  {journal} {\bibinfo  {journal} {BMC Syst. Biol.}\
  }\textbf {\bibinfo {volume} {6}},\ \bibinfo {pages} {39} (\bibinfo {year}
  {2012})}\BibitemShut {NoStop}%
\bibitem [{\citenamefont {McAdams}\ and\ \citenamefont
  {Arkin}(1999)}]{McAdams:1999ex}%
  \BibitemOpen
  \bibfield  {author} {\bibinfo {author} {\bibfnamefont {H.~H.}\ \bibnamefont
  {McAdams}}\ and\ \bibinfo {author} {\bibfnamefont {A.}~\bibnamefont
  {Arkin}},\ }\href@noop {} {\bibfield  {journal} {\bibinfo  {journal} {Trends
  Genet.}\ }\textbf {\bibinfo {volume} {15}},\ \bibinfo {pages} {65} (\bibinfo
  {year} {1999})}\BibitemShut {NoStop}%
\bibitem [{\citenamefont {Ozbudak}\ \emph {et~al.}(2004)\citenamefont
  {Ozbudak}, \citenamefont {Thattai}, \citenamefont {Lim}, \citenamefont
  {Shraiman},\ and\ \citenamefont {van Oudenaarden}}]{Ozbudak:2004gi}%
  \BibitemOpen
  \bibfield  {author} {\bibinfo {author} {\bibfnamefont {E.~M.}\ \bibnamefont
  {Ozbudak}}, \bibinfo {author} {\bibfnamefont {M.}~\bibnamefont {Thattai}},
  \bibinfo {author} {\bibfnamefont {H.~N.}\ \bibnamefont {Lim}}, \bibinfo
  {author} {\bibfnamefont {B.~I.}\ \bibnamefont {Shraiman}}, \ and\ \bibinfo
  {author} {\bibfnamefont {A.}~\bibnamefont {van Oudenaarden}},\ }\href@noop {}
  {\bibfield  {journal} {\bibinfo  {journal} {Nature}\ }\textbf {\bibinfo
  {volume} {427}},\ \bibinfo {pages} {737} (\bibinfo {year}
  {2004})}\BibitemShut {NoStop}%
\bibitem [{\citenamefont {Veening}, \citenamefont {Smits},\ and\ \citenamefont
  {Kuipers}(2008)}]{Veening:2013bj}%
  \BibitemOpen
  \bibfield  {author} {\bibinfo {author} {\bibfnamefont {J.-W.}\ \bibnamefont
  {Veening}}, \bibinfo {author} {\bibfnamefont {W.~K.}\ \bibnamefont {Smits}},
  \ and\ \bibinfo {author} {\bibfnamefont {O.~P.}\ \bibnamefont {Kuipers}},\
  }\href@noop {} {\bibfield  {journal} {\bibinfo  {journal} {Annu. Rev.
  Microbiol.}\ }\textbf {\bibinfo {volume} {62}},\ \bibinfo {pages} {193}
  (\bibinfo {year} {2008})}\BibitemShut {NoStop}%
\bibitem [{\citenamefont {Horsthemke}\ and\ \citenamefont
  {Lefever}(1984)}]{Horsthemke:1984ut}%
  \BibitemOpen
  \bibfield  {author} {\bibinfo {author} {\bibfnamefont {W.}~\bibnamefont
  {Horsthemke}}\ and\ \bibinfo {author} {\bibfnamefont {R.}~\bibnamefont
  {Lefever}},\ }\href@noop {} {\emph {\bibinfo {title} {{Noise-induced
  transitions: Theory and applications in physics, chemistry, and biology}}}},\
  \bibinfo {edition} {1st}\ ed.,\ Vol.~\bibinfo {volume} {15}\ (\bibinfo
  {publisher} {Springer-Verlag},\ \bibinfo {address} {Berlin and New York},\
  \bibinfo {year} {1984})\BibitemShut {NoStop}%
\bibitem [{\citenamefont {Wang}\ \emph {et~al.}(2012)\citenamefont {Wang},
  \citenamefont {Dearing}, \citenamefont {Langdon}, \citenamefont {Zhang},
  \citenamefont {Yang}, \citenamefont {Dakos},\ and\ \citenamefont
  {Scheffer}}]{Wang:2012cd}%
  \BibitemOpen
  \bibfield  {author} {\bibinfo {author} {\bibfnamefont {R.}~\bibnamefont
  {Wang}}, \bibinfo {author} {\bibfnamefont {J.~A.}\ \bibnamefont {Dearing}},
  \bibinfo {author} {\bibfnamefont {P.~G.}\ \bibnamefont {Langdon}}, \bibinfo
  {author} {\bibfnamefont {E.}~\bibnamefont {Zhang}}, \bibinfo {author}
  {\bibfnamefont {X.}~\bibnamefont {Yang}}, \bibinfo {author} {\bibfnamefont
  {V.}~\bibnamefont {Dakos}}, \ and\ \bibinfo {author} {\bibfnamefont
  {M.}~\bibnamefont {Scheffer}},\ }\href@noop {} {\bibfield  {journal}
  {\bibinfo  {journal} {Nature}\ }\textbf {\bibinfo {volume} {492}},\ \bibinfo
  {pages} {419} (\bibinfo {year} {2012})}\BibitemShut {NoStop}%
\bibitem [{\citenamefont {E}, \citenamefont {Liu},\ and\ \citenamefont
  {Vanden-Eijnden}(2005)}]{E:2005bq}%
  \BibitemOpen
  \bibfield  {author} {\bibinfo {author} {\bibfnamefont {W.}~\bibnamefont {E}},
  \bibinfo {author} {\bibfnamefont {D.}~\bibnamefont {Liu}}, \ and\ \bibinfo
  {author} {\bibfnamefont {E.}~\bibnamefont {Vanden-Eijnden}},\ }\href@noop {}
  {\bibfield  {journal} {\bibinfo  {journal} {Comm. Pure Appl. Math.}\ }\textbf
  {\bibinfo {volume} {58}},\ \bibinfo {pages} {1544} (\bibinfo {year}
  {2005})}\BibitemShut {NoStop}%
\bibitem [{\citenamefont {Staver}, \citenamefont {Archibald},\ and\
  \citenamefont {Levin}(2011)}]{Staver:2011bz}%
  \BibitemOpen
  \bibfield  {author} {\bibinfo {author} {\bibfnamefont {A.~C.}\ \bibnamefont
  {Staver}}, \bibinfo {author} {\bibfnamefont {S.}~\bibnamefont {Archibald}}, \
  and\ \bibinfo {author} {\bibfnamefont {S.~A.}\ \bibnamefont {Levin}},\
  }\href@noop {} {\bibfield  {journal} {\bibinfo  {journal} {Science}\ }\textbf
  {\bibinfo {volume} {334}},\ \bibinfo {pages} {230} (\bibinfo {year}
  {2011})}\BibitemShut {NoStop}%
\bibitem [{\citenamefont {Scheffer}\ \emph {et~al.}(2012)\citenamefont
  {Scheffer}, \citenamefont {Carpenter}, \citenamefont {Lenton}, \citenamefont
  {Bascompte}, \citenamefont {Brock}, \citenamefont {Dakos}, \citenamefont
  {van~de Koppel}, \citenamefont {van~de Leemput}, \citenamefont {Levin},
  \citenamefont {van Nes}, \citenamefont {Pascual},\ and\ \citenamefont
  {Vandermeer}}]{Scheffer:2012ct}%
  \BibitemOpen
  \bibfield  {author} {\bibinfo {author} {\bibfnamefont {M.}~\bibnamefont
  {Scheffer}}, \bibinfo {author} {\bibfnamefont {S.~R.}\ \bibnamefont
  {Carpenter}}, \bibinfo {author} {\bibfnamefont {T.~M.}\ \bibnamefont
  {Lenton}}, \bibinfo {author} {\bibfnamefont {J.}~\bibnamefont {Bascompte}},
  \bibinfo {author} {\bibfnamefont {W.}~\bibnamefont {Brock}}, \bibinfo
  {author} {\bibfnamefont {V.}~\bibnamefont {Dakos}}, \bibinfo {author}
  {\bibfnamefont {J.}~\bibnamefont {van~de Koppel}}, \bibinfo {author}
  {\bibfnamefont {I.~A.}\ \bibnamefont {van~de Leemput}}, \bibinfo {author}
  {\bibfnamefont {S.~A.}\ \bibnamefont {Levin}}, \bibinfo {author}
  {\bibfnamefont {E.~H.}\ \bibnamefont {van Nes}}, \bibinfo {author}
  {\bibfnamefont {M.}~\bibnamefont {Pascual}}, \ and\ \bibinfo {author}
  {\bibfnamefont {J.}~\bibnamefont {Vandermeer}},\ }\href@noop {} {\bibfield
  {journal} {\bibinfo  {journal} {Science}\ }\textbf {\bibinfo {volume}
  {338}},\ \bibinfo {pages} {344} (\bibinfo {year} {2012})}\BibitemShut
  {NoStop}%
\bibitem [{\citenamefont {Fenichel}(1979)}]{Fenichel:1979dz}%
  \BibitemOpen
  \bibfield  {author} {\bibinfo {author} {\bibfnamefont {N.}~\bibnamefont
  {Fenichel}},\ }\href@noop {} {\bibfield  {journal} {\bibinfo  {journal} {J.
  Differ. Equations}\ }\textbf {\bibinfo {volume} {31}},\ \bibinfo {pages} {53}
  (\bibinfo {year} {1979})}\BibitemShut {NoStop}%
\bibitem [{\citenamefont {Desroches}\ \emph {et~al.}(2012)\citenamefont
  {Desroches}, \citenamefont {Guckenheimer}, \citenamefont {Krauskopf},
  \citenamefont {Kuehn}, \citenamefont {Osinga},\ and\ \citenamefont
  {Wechselberger}}]{Desroches:2012cj}%
  \BibitemOpen
  \bibfield  {author} {\bibinfo {author} {\bibfnamefont {M.}~\bibnamefont
  {Desroches}}, \bibinfo {author} {\bibfnamefont {J.}~\bibnamefont
  {Guckenheimer}}, \bibinfo {author} {\bibfnamefont {B.}~\bibnamefont
  {Krauskopf}}, \bibinfo {author} {\bibfnamefont {C.}~\bibnamefont {Kuehn}},
  \bibinfo {author} {\bibfnamefont {H.~M.}\ \bibnamefont {Osinga}}, \ and\
  \bibinfo {author} {\bibfnamefont {M.}~\bibnamefont {Wechselberger}},\
  }\href@noop {} {\bibfield  {journal} {\bibinfo  {journal} {SIAM Rev.}\
  }\textbf {\bibinfo {volume} {54}},\ \bibinfo {pages} {211} (\bibinfo {year}
  {2012})}\BibitemShut {NoStop}%
\bibitem [{\citenamefont {Boxler}(1989)}]{Boxler:1989cb}%
  \BibitemOpen
  \bibfield  {author} {\bibinfo {author} {\bibfnamefont {P.}~\bibnamefont
  {Boxler}},\ }\href@noop {} {\bibfield  {journal} {\bibinfo  {journal}
  {Probab. Theory Rel.}\ }\textbf {\bibinfo {volume} {83}},\ \bibinfo {pages}
  {509} (\bibinfo {year} {1989})}\BibitemShut {NoStop}%
\bibitem [{\citenamefont {Constable}, \citenamefont {McKane},\ and\
  \citenamefont {Rogers}(2013)}]{Constable:2013fu}%
  \BibitemOpen
  \bibfield  {author} {\bibinfo {author} {\bibfnamefont {G.~W.~A.}\
  \bibnamefont {Constable}}, \bibinfo {author} {\bibfnamefont {A.~J.}\
  \bibnamefont {McKane}}, \ and\ \bibinfo {author} {\bibfnamefont
  {T.}~\bibnamefont {Rogers}},\ }\href@noop {} {\bibfield  {journal} {\bibinfo
  {journal} {J. Phys. A: Math. Theor.}\ }\textbf {\bibinfo {volume} {46}},\
  \bibinfo {pages} {295002} (\bibinfo {year} {2013})}\BibitemShut {NoStop}%
\bibitem [{\citenamefont {Wainrib}(2013)}]{Wainrib:2012vj}%
  \BibitemOpen
  \bibfield  {author} {\bibinfo {author} {\bibfnamefont {G.}~\bibnamefont
  {Wainrib}},\ }\href@noop {} {\bibfield  {journal} {\bibinfo  {journal}
  {Electronic Communications in Probability}\ }\textbf {\bibinfo {volume}
  {18}},\ \bibinfo {pages} {1} (\bibinfo {year} {2013})}\BibitemShut {NoStop}%
\bibitem [{\citenamefont {Gardiner}(2004)}]{gardiner2004handbook}%
  \BibitemOpen
  \bibfield  {author} {\bibinfo {author} {\bibfnamefont {C.~W.}\ \bibnamefont
  {Gardiner}},\ }\href@noop {} {\emph {\bibinfo {title} {{Handbook of
  Stochastic Methods for Physics, Chemistry and the Natural Sciences}}}},\
  \bibinfo {edition} {3rd}\ ed.\ (\bibinfo  {publisher} {Springer-Verlag, New
  York},\ \bibinfo {year} {2004})\BibitemShut {NoStop}%
\bibitem [{\citenamefont {Cao}, \citenamefont {Gillespie},\ and\ \citenamefont
  {Petzold}(2005)}]{Cao:2005gj}%
  \BibitemOpen
  \bibfield  {author} {\bibinfo {author} {\bibfnamefont {Y.}~\bibnamefont
  {Cao}}, \bibinfo {author} {\bibfnamefont {D.~T.}\ \bibnamefont {Gillespie}},
  \ and\ \bibinfo {author} {\bibfnamefont {L.~R.}\ \bibnamefont {Petzold}},\
  }\href@noop {} {\bibfield  {journal} {\bibinfo  {journal} {J. Chem. Phys.}\
  }\textbf {\bibinfo {volume} {122}},\ \bibinfo {pages} {014116} (\bibinfo
  {year} {2005})}\BibitemShut {NoStop}%
\bibitem [{\citenamefont {Rao}\ and\ \citenamefont {Arkin}(2003)}]{Rao:2003gu}%
  \BibitemOpen
  \bibfield  {author} {\bibinfo {author} {\bibfnamefont {C.~V.}\ \bibnamefont
  {Rao}}\ and\ \bibinfo {author} {\bibfnamefont {A.~P.}\ \bibnamefont
  {Arkin}},\ }\href@noop {} {\bibfield  {journal} {\bibinfo  {journal} {J.
  Chem. Phys.}\ }\textbf {\bibinfo {volume} {118}},\ \bibinfo {pages} {4999}
  (\bibinfo {year} {2003})}\BibitemShut {NoStop}%
\bibitem [{\citenamefont {Newby}\ and\ \citenamefont
  {Chapman}(2013)}]{Newby:2013gx}%
  \BibitemOpen
  \bibfield  {author} {\bibinfo {author} {\bibfnamefont {J.~M.}\ \bibnamefont
  {Newby}}\ and\ \bibinfo {author} {\bibfnamefont {S.~J.}\ \bibnamefont
  {Chapman}},\ }\href@noop {} {\bibfield  {journal} {\bibinfo  {journal} {J.
  Math. Biol.}\ } (\bibinfo {year} {2013})}\BibitemShut {NoStop}%
\bibitem [{\citenamefont {Ward}(1998)}]{Ward:1998ve}%
  \BibitemOpen
  \bibfield  {author} {\bibinfo {author} {\bibfnamefont {M.~J.}\ \bibnamefont
  {Ward}},\ }in\ \href@noop {} {\emph {\bibinfo {booktitle} {Analyzing
  Multiscale Phenomena Using Singular Perturbation Methods}}},\ \bibinfo
  {editor} {edited by\ \bibinfo {editor} {\bibfnamefont {J.}~\bibnamefont
  {Cronin}}\ and\ \bibinfo {editor} {\bibfnamefont {R.}~\bibnamefont
  {O'Malley}}}\ (\bibinfo  {publisher} {AMS publications},\ \bibinfo {address}
  {Providence, RI},\ \bibinfo {year} {1998})\ pp.\ \bibinfo {pages}
  {151--184}\BibitemShut {NoStop}%
\bibitem [{\citenamefont {Hinch}\ and\ \citenamefont
  {Chapman}(2005)}]{Hinch:2005eg}%
  \BibitemOpen
  \bibfield  {author} {\bibinfo {author} {\bibfnamefont {R.}~\bibnamefont
  {Hinch}}\ and\ \bibinfo {author} {\bibfnamefont {S.~J.}\ \bibnamefont
  {Chapman}},\ }\href@noop {} {\bibfield  {journal} {\bibinfo  {journal} {Eur.
  J. Appl. Math}\ }\textbf {\bibinfo {volume} {16}},\ \bibinfo {pages} {427}
  (\bibinfo {year} {2005})}\BibitemShut {NoStop}%
\bibitem [{\citenamefont {Newby}(2012)}]{Newby:2012fz}%
  \BibitemOpen
  \bibfield  {author} {\bibinfo {author} {\bibfnamefont {J.~M.}\ \bibnamefont
  {Newby}},\ }\href@noop {} {\bibfield  {journal} {\bibinfo  {journal} {Phys.
  Biol.}\ }\textbf {\bibinfo {volume} {9}},\ \bibinfo {pages} {026002}
  (\bibinfo {year} {2012})}\BibitemShut {NoStop}%
\bibitem [{\citenamefont {Hartmann}\ \emph {et~al.}(2013)\citenamefont
  {Hartmann}, \citenamefont {Banisch}, \citenamefont {Sarich}, \citenamefont
  {Badowski},\ and\ \citenamefont {Sch{\"u}tte}}]{Hartmann:2013wi}%
  \BibitemOpen
  \bibfield  {author} {\bibinfo {author} {\bibfnamefont {C.}~\bibnamefont
  {Hartmann}}, \bibinfo {author} {\bibfnamefont {R.}~\bibnamefont {Banisch}},
  \bibinfo {author} {\bibfnamefont {M.}~\bibnamefont {Sarich}}, \bibinfo
  {author} {\bibfnamefont {T.}~\bibnamefont {Badowski}}, \ and\ \bibinfo
  {author} {\bibfnamefont {C.}~\bibnamefont {Sch{\"u}tte}},\ }\href@noop {}
  {\bibfield  {journal} {\bibinfo  {journal} {Entropy}\ }\textbf {\bibinfo
  {volume} {16}},\ \bibinfo {pages} {350} (\bibinfo {year} {2013})}\BibitemShut
  {NoStop}%
\bibitem [{\citenamefont {Schl{\"o}gl}(1972)}]{Schlogl:1972he}%
  \BibitemOpen
  \bibfield  {author} {\bibinfo {author} {\bibfnamefont {F.}~\bibnamefont
  {Schl{\"o}gl}},\ }\href@noop {} {\bibfield  {journal} {\bibinfo  {journal}
  {Z. Phys.}\ }\textbf {\bibinfo {volume} {253}},\ \bibinfo {pages} {147}
  (\bibinfo {year} {1972})}\BibitemShut {NoStop}%
\bibitem [{\citenamefont {Erban}\ \emph {et~al.}(2009)\citenamefont {Erban},
  \citenamefont {Chapman}, \citenamefont {Kevrekidis},\ and\ \citenamefont
  {Vejchodsk{\'y}}}]{Erban:2009ew}%
  \BibitemOpen
  \bibfield  {author} {\bibinfo {author} {\bibfnamefont {R.}~\bibnamefont
  {Erban}}, \bibinfo {author} {\bibfnamefont {S.~J.}\ \bibnamefont {Chapman}},
  \bibinfo {author} {\bibfnamefont {I.~G.}\ \bibnamefont {Kevrekidis}}, \ and\
  \bibinfo {author} {\bibfnamefont {T.}~\bibnamefont {Vejchodsk{\'y}}},\
  }\href@noop {} {\bibfield  {journal} {\bibinfo  {journal} {SIAM J. Appl.
  Math.}\ }\textbf {\bibinfo {volume} {70}},\ \bibinfo {pages} {984} (\bibinfo
  {year} {2009})}\BibitemShut {NoStop}%
\bibitem [{Note1()}]{Note1}%
  \BibitemOpen
  \bibinfo {note} {Some authors would use the convention that the propensity
  function for the third reaction, $2 X \mathrel \bgroup \protect \stack@relbin
  []{k_3/\epsilon }{\DOTSB \protect \relbar \protect \joinrel \rightarrow }
  3X$, is $k_3 x (x-1)/2\epsilon $ instead of $k_3 x (x-1)/\epsilon $. See
  Ref.~\protect \rev@citealpnum {Gillespie:1977dc} for a discussion on
  conventions regarding reaction rates.}\BibitemShut {Stop}%
\bibitem [{\citenamefont {Gillespie}(1977)}]{Gillespie:1977dc}%
  \BibitemOpen
  \bibfield  {author} {\bibinfo {author} {\bibfnamefont {D.~T.}\ \bibnamefont
  {Gillespie}},\ }\href@noop {} {\bibfield  {journal} {\bibinfo  {journal} {J.
  Phys. Chem.}\ }\textbf {\bibinfo {volume} {81}},\ \bibinfo {pages} {2340}
  (\bibinfo {year} {1977})}\BibitemShut {NoStop}%
\bibitem [{\citenamefont {Gibson}\ and\ \citenamefont
  {Bruck}(2000)}]{Gibson:2000jq}%
  \BibitemOpen
  \bibfield  {author} {\bibinfo {author} {\bibfnamefont {M.~A.}\ \bibnamefont
  {Gibson}}\ and\ \bibinfo {author} {\bibfnamefont {J.}~\bibnamefont {Bruck}},\
  }\href@noop {} {\bibfield  {journal} {\bibinfo  {journal} {J. Phys. Chem. A}\
  }\textbf {\bibinfo {volume} {104}},\ \bibinfo {pages} {1876} (\bibinfo {year}
  {2000})}\BibitemShut {NoStop}%
\bibitem [{\citenamefont {Cao}, \citenamefont {Gillespie},\ and\ \citenamefont
  {Petzold}(2006)}]{Cao:2006hq}%
  \BibitemOpen
  \bibfield  {author} {\bibinfo {author} {\bibfnamefont {Y.}~\bibnamefont
  {Cao}}, \bibinfo {author} {\bibfnamefont {D.~T.}\ \bibnamefont {Gillespie}},
  \ and\ \bibinfo {author} {\bibfnamefont {L.~R.}\ \bibnamefont {Petzold}},\
  }\href@noop {} {\bibfield  {journal} {\bibinfo  {journal} {J. Chem. Phys.}\
  }\textbf {\bibinfo {volume} {124}},\ \bibinfo {pages} {044109} (\bibinfo
  {year} {2006})}\BibitemShut {NoStop}%
\bibitem [{\citenamefont {Gillespie}(2000)}]{Gillespie:2000ig}%
  \BibitemOpen
  \bibfield  {author} {\bibinfo {author} {\bibfnamefont {D.~T.}\ \bibnamefont
  {Gillespie}},\ }\href@noop {} {\bibfield  {journal} {\bibinfo  {journal} {J.
  Chem. Phys.}\ }\textbf {\bibinfo {volume} {113}},\ \bibinfo {pages} {297}
  (\bibinfo {year} {2000})}\BibitemShut {NoStop}%
\bibitem [{Note2()}]{Note2}%
  \BibitemOpen
  \bibinfo {note} {This is in contrast to the Parallel Replica Algorithm,\cite
  {LeBris:2012et} which defines quasi-stationary distributions using absorbing
  boundary conditions.}\BibitemShut {Stop}%
\bibitem [{\citenamefont {Risken}(1996)}]{Risken:1996vl}%
  \BibitemOpen
  \bibfield  {author} {\bibinfo {author} {\bibfnamefont {H.}~\bibnamefont
  {Risken}},\ }\href@noop {} {\emph {\bibinfo {title} {{The Fokker-Planck
  Equation}}}},\ Methods of Solution and Applications\ (\bibinfo  {publisher}
  {Springer},\ \bibinfo {year} {1996})\BibitemShut {NoStop}%
\bibitem [{\citenamefont {Le~Bris}\ \emph {et~al.}(2012)\citenamefont
  {Le~Bris}, \citenamefont {Leli{\`e}vre}, \citenamefont {Luskin},\ and\
  \citenamefont {Perez}}]{LeBris:2012et}%
  \BibitemOpen
  \bibfield  {author} {\bibinfo {author} {\bibfnamefont {C.}~\bibnamefont
  {Le~Bris}}, \bibinfo {author} {\bibfnamefont {T.}~\bibnamefont
  {Leli{\`e}vre}}, \bibinfo {author} {\bibfnamefont {M.}~\bibnamefont
  {Luskin}}, \ and\ \bibinfo {author} {\bibfnamefont {D.}~\bibnamefont
  {Perez}},\ }\href@noop {} {\bibfield  {journal} {\bibinfo  {journal} {Monte
  Carlo Methods Appl.}\ }\textbf {\bibinfo {volume} {18}},\ \bibinfo {pages}
  {119} (\bibinfo {year} {2012})}\BibitemShut {NoStop}%
\bibitem [{\citenamefont {Ovaskainen}\ and\ \citenamefont
  {Meerson}(2010)}]{Ovaskainen:2010wm}%
  \BibitemOpen
  \bibfield  {author} {\bibinfo {author} {\bibfnamefont {O.}~\bibnamefont
  {Ovaskainen}}\ and\ \bibinfo {author} {\bibfnamefont {B.}~\bibnamefont
  {Meerson}},\ }\href@noop {} {\bibfield  {journal} {\bibinfo  {journal}
  {Trends Ecol. Evol.}\ }\textbf {\bibinfo {volume} {25}},\ \bibinfo {pages}
  {643} (\bibinfo {year} {2010})}\BibitemShut {NoStop}%
\bibitem [{\citenamefont {Brassil}(2001)}]{Brassil:2001eg}%
  \BibitemOpen
  \bibfield  {author} {\bibinfo {author} {\bibfnamefont {C.~E.}\ \bibnamefont
  {Brassil}},\ }\href@noop {} {\bibfield  {journal} {\bibinfo  {journal} {Ecol.
  model.}\ }\textbf {\bibinfo {volume} {143}},\ \bibinfo {pages} {9} (\bibinfo
  {year} {2001})}\BibitemShut {NoStop}%
\bibitem [{\citenamefont {Dennis}(2002)}]{Dennis:2002iy}%
  \BibitemOpen
  \bibfield  {author} {\bibinfo {author} {\bibfnamefont {B.}~\bibnamefont
  {Dennis}},\ }\href@noop {} {\bibfield  {journal} {\bibinfo  {journal}
  {Oikos}\ }\textbf {\bibinfo {volume} {96}},\ \bibinfo {pages} {389} (\bibinfo
  {year} {2002})}\BibitemShut {NoStop}%
\bibitem [{\citenamefont {Schreiber}(2003)}]{Schreiber:2003hf}%
  \BibitemOpen
  \bibfield  {author} {\bibinfo {author} {\bibfnamefont {S.~J.}\ \bibnamefont
  {Schreiber}},\ }\href@noop {} {\bibfield  {journal} {\bibinfo  {journal}
  {Theor. Popul. Biol.}\ }\textbf {\bibinfo {volume} {64}},\ \bibinfo {pages}
  {201} (\bibinfo {year} {2003})}\BibitemShut {NoStop}%
\bibitem [{\citenamefont {Palamara}\ \emph {et~al.}(2013)\citenamefont
  {Palamara}, \citenamefont {Delius}, \citenamefont {Smith},\ and\
  \citenamefont {Petchey}}]{Palamara:2012uoa}%
  \BibitemOpen
  \bibfield  {author} {\bibinfo {author} {\bibfnamefont {G.~M.}\ \bibnamefont
  {Palamara}}, \bibinfo {author} {\bibfnamefont {G.~W.}\ \bibnamefont
  {Delius}}, \bibinfo {author} {\bibfnamefont {M.~J.}\ \bibnamefont {Smith}}, \
  and\ \bibinfo {author} {\bibfnamefont {O.~L.}\ \bibnamefont {Petchey}},\
  }\href@noop {} {\bibfield  {journal} {\bibinfo  {journal} {J. Theor. Biol.}\
  }\textbf {\bibinfo {volume} {334}},\ \bibinfo {pages} {61} (\bibinfo {year}
  {2013})}\BibitemShut {NoStop}%
\bibitem [{\citenamefont {N{\aa}sell}(2001)}]{Nasell:2001uc}%
  \BibitemOpen
  \bibfield  {author} {\bibinfo {author} {\bibfnamefont {I.}~\bibnamefont
  {N{\aa}sell}},\ }\href@noop {} {\bibfield  {journal} {\bibinfo  {journal} {J.
  Theor. Biol.}\ }\textbf {\bibinfo {volume} {211}},\ \bibinfo {pages} {11}
  (\bibinfo {year} {2001})}\BibitemShut {NoStop}%
\bibitem [{\citenamefont {Turchin}(2003)}]{Turchin:2003tb}%
  \BibitemOpen
  \bibfield  {author} {\bibinfo {author} {\bibfnamefont {P.}~\bibnamefont
  {Turchin}},\ }\href@noop {} {\emph {\bibinfo {title} {{Complex Population
  Dynamics: A Theoretical/Empirical Synthesis}}}}\ (\bibinfo  {publisher}
  {Princeton University Press},\ \bibinfo {year} {2003})\BibitemShut {NoStop}%
\bibitem [{\citenamefont {Grasman}\ and\ \citenamefont
  {HilleRisLambers}(1997)}]{Grasman:1997vp}%
  \BibitemOpen
  \bibfield  {author} {\bibinfo {author} {\bibfnamefont {J.}~\bibnamefont
  {Grasman}}\ and\ \bibinfo {author} {\bibfnamefont {R.}~\bibnamefont
  {HilleRisLambers}},\ }\href@noop {} {\bibfield  {journal} {\bibinfo
  {journal} {Ecol. model.}\ }\textbf {\bibinfo {volume} {103}},\ \bibinfo
  {pages} {71} (\bibinfo {year} {1997})}\BibitemShut {NoStop}%
\bibitem [{Note3()}]{Note3}%
  \BibitemOpen
  \bibinfo {note} {This is because if we waited long enough in the stochastic
  model, extinction would eventually occur, and therefore the stationary
  density is not defined. In other words, there is a ``leak'' at
  $x=0$.}\BibitemShut {Stop}%
\bibitem [{\citenamefont {Kessler}\ and\ \citenamefont
  {Shnerb}(2007)}]{Kessler:2007ex}%
  \BibitemOpen
  \bibfield  {author} {\bibinfo {author} {\bibfnamefont {D.~A.}\ \bibnamefont
  {Kessler}}\ and\ \bibinfo {author} {\bibfnamefont {N.~M.}\ \bibnamefont
  {Shnerb}},\ }\href@noop {} {\bibfield  {journal} {\bibinfo  {journal} {J.
  Stat. Phys.}\ }\textbf {\bibinfo {volume} {127}},\ \bibinfo {pages} {861}
  (\bibinfo {year} {2007})}\BibitemShut {NoStop}%
\bibitem [{\citenamefont {Assaf}\ and\ \citenamefont
  {Meerson}(2010)}]{Assaf:2010de}%
  \BibitemOpen
  \bibfield  {author} {\bibinfo {author} {\bibfnamefont {M.}~\bibnamefont
  {Assaf}}\ and\ \bibinfo {author} {\bibfnamefont {B.}~\bibnamefont
  {Meerson}},\ }\href@noop {} {\bibfield  {journal} {\bibinfo  {journal} {Phys.
  Rev. E}\ }\textbf {\bibinfo {volume} {81}},\ \bibinfo {pages} {021116}
  (\bibinfo {year} {2010})}\BibitemShut {NoStop}%
\bibitem [{\citenamefont {Prinz}\ \emph {et~al.}(2011)\citenamefont {Prinz},
  \citenamefont {Wu}, \citenamefont {Sarich}, \citenamefont {Keller},
  \citenamefont {Senne}, \citenamefont {Held}, \citenamefont {Chodera},
  \citenamefont {Sch{\"u}tte},\ and\ \citenamefont {No{\'e}}}]{Prinz:2011id}%
  \BibitemOpen
  \bibfield  {author} {\bibinfo {author} {\bibfnamefont {J.-H.}\ \bibnamefont
  {Prinz}}, \bibinfo {author} {\bibfnamefont {H.}~\bibnamefont {Wu}}, \bibinfo
  {author} {\bibfnamefont {M.}~\bibnamefont {Sarich}}, \bibinfo {author}
  {\bibfnamefont {B.}~\bibnamefont {Keller}}, \bibinfo {author} {\bibfnamefont
  {M.}~\bibnamefont {Senne}}, \bibinfo {author} {\bibfnamefont
  {M.}~\bibnamefont {Held}}, \bibinfo {author} {\bibfnamefont {J.~D.}\
  \bibnamefont {Chodera}}, \bibinfo {author} {\bibfnamefont {C.}~\bibnamefont
  {Sch{\"u}tte}}, \ and\ \bibinfo {author} {\bibfnamefont {F.}~\bibnamefont
  {No{\'e}}},\ }\href@noop {} {\bibfield  {journal} {\bibinfo  {journal} {J.
  Chem. Phys.}\ }\textbf {\bibinfo {volume} {134}},\ \bibinfo {pages} {174105}
  (\bibinfo {year} {2011})}\BibitemShut {NoStop}%
\bibitem [{\citenamefont {Abad}, \citenamefont {Reingruber},\ and\
  \citenamefont {Sansom}(2009)}]{Abad:2009fd}%
  \BibitemOpen
  \bibfield  {author} {\bibinfo {author} {\bibfnamefont {E.}~\bibnamefont
  {Abad}}, \bibinfo {author} {\bibfnamefont {J.}~\bibnamefont {Reingruber}}, \
  and\ \bibinfo {author} {\bibfnamefont {M.~S.~P.}\ \bibnamefont {Sansom}},\
  }\href@noop {} {\bibfield  {journal} {\bibinfo  {journal} {J. Chem. Phys.}\
  }\textbf {\bibinfo {volume} {130}},\ \bibinfo {pages} {085101} (\bibinfo
  {year} {2009})}\BibitemShut {NoStop}%
\bibitem [{\citenamefont {Chen}, \citenamefont {Erban},\ and\ \citenamefont
  {Chapman}(2014)}]{Chen:2014wh}%
  \BibitemOpen
  \bibfield  {author} {\bibinfo {author} {\bibfnamefont {W.}~\bibnamefont
  {Chen}}, \bibinfo {author} {\bibfnamefont {R.}~\bibnamefont {Erban}}, \ and\
  \bibinfo {author} {\bibfnamefont {S.~J.}\ \bibnamefont {Chapman}},\
  }\href@noop {} {\bibfield  {journal} {\bibinfo  {journal} {SIAM J. Appl.
  Math.}\ }\textbf {\bibinfo {volume} {74}},\ \bibinfo {pages} {208} (\bibinfo
  {year} {2014})}\BibitemShut {NoStop}%
\bibitem [{\citenamefont {Bressloff}\ and\ \citenamefont
  {Newby}(2013{\natexlab{a}})}]{Bressloff:2013ve}%
  \BibitemOpen
  \bibfield  {author} {\bibinfo {author} {\bibfnamefont {P.~C.}\ \bibnamefont
  {Bressloff}}\ and\ \bibinfo {author} {\bibfnamefont {J.~M.}\ \bibnamefont
  {Newby}},\ }\href@noop {} {\bibfield  {journal} {\bibinfo  {journal} {SIAM J.
  Appl. Dyn. Syst.}\ }\textbf {\bibinfo {volume} {12}},\ \bibinfo {pages}
  {1394} (\bibinfo {year} {2013}{\natexlab{a}})}\BibitemShut {NoStop}%
\bibitem [{\citenamefont {Dakos}\ \emph {et~al.}(2012)\citenamefont {Dakos},
  \citenamefont {Carpenter}, \citenamefont {Brock}, \citenamefont {Ellison},
  \citenamefont {Guttal}, \citenamefont {Ives}, \citenamefont {K{\'e}fi},
  \citenamefont {Livina}, \citenamefont {Seekell}, \citenamefont {van Nes},\
  and\ \citenamefont {Scheffer}}]{Dakos:2012ft}%
  \BibitemOpen
  \bibfield  {author} {\bibinfo {author} {\bibfnamefont {V.}~\bibnamefont
  {Dakos}}, \bibinfo {author} {\bibfnamefont {S.~R.}\ \bibnamefont
  {Carpenter}}, \bibinfo {author} {\bibfnamefont {W.~A.}\ \bibnamefont
  {Brock}}, \bibinfo {author} {\bibfnamefont {A.~M.}\ \bibnamefont {Ellison}},
  \bibinfo {author} {\bibfnamefont {V.}~\bibnamefont {Guttal}}, \bibinfo
  {author} {\bibfnamefont {A.~R.}\ \bibnamefont {Ives}}, \bibinfo {author}
  {\bibfnamefont {S.}~\bibnamefont {K{\'e}fi}}, \bibinfo {author}
  {\bibfnamefont {V.}~\bibnamefont {Livina}}, \bibinfo {author} {\bibfnamefont
  {D.~A.}\ \bibnamefont {Seekell}}, \bibinfo {author} {\bibfnamefont {E.~H.}\
  \bibnamefont {van Nes}}, \ and\ \bibinfo {author} {\bibfnamefont
  {M.}~\bibnamefont {Scheffer}},\ }\href@noop {} {\bibfield  {journal}
  {\bibinfo  {journal} {PLoS ONE}\ }\textbf {\bibinfo {volume} {7}},\ \bibinfo
  {pages} {e41010} (\bibinfo {year} {2012})}\BibitemShut {NoStop}%
\bibitem [{\citenamefont {Geissbuehler}\ and\ \citenamefont
  {Lasser}(2013)}]{Geissbuehler:2013er}%
  \BibitemOpen
  \bibfield  {author} {\bibinfo {author} {\bibfnamefont {M.}~\bibnamefont
  {Geissbuehler}}\ and\ \bibinfo {author} {\bibfnamefont {T.}~\bibnamefont
  {Lasser}},\ }\href@noop {} {\bibfield  {journal} {\bibinfo  {journal} {Opt.
  Express}\ }\textbf {\bibinfo {volume} {21}},\ \bibinfo {pages} {9862}
  (\bibinfo {year} {2013})}\BibitemShut {NoStop}%
\bibitem [{\citenamefont {Bressloff}\ and\ \citenamefont
  {Newby}(2013{\natexlab{b}})}]{bressloff:2013wn}%
  \BibitemOpen
  \bibfield  {author} {\bibinfo {author} {\bibfnamefont {P.~C.}\ \bibnamefont
  {Bressloff}}\ and\ \bibinfo {author} {\bibfnamefont {J.~M.}\ \bibnamefont
  {Newby}},\ }in\ \href@noop {} {\emph {\bibinfo {booktitle} {First-Passage
  Phenomena and Their Applications}}},\ \bibinfo {editor} {edited by\ \bibinfo
  {editor} {\bibfnamefont {R.}~\bibnamefont {Metzler}}, \bibinfo {editor}
  {\bibfnamefont {G.}~\bibnamefont {Oshanin}}, \ and\ \bibinfo {editor}
  {\bibfnamefont {S.}~\bibnamefont {Redner}}}\ (\bibinfo  {publisher} {World
  Scientific},\ \bibinfo {year} {2013})\ pp.\ \bibinfo {pages}
  {1--29}\BibitemShut {NoStop}%
\end{thebibliography}

%

\end{document}